\documentclass[fleqn,usenatbib]{mnras}
\usepackage{newtxtext,newtxmath}
\usepackage[T1]{fontenc}
\usepackage{ae,aecompl}

\usepackage{graphicx}	% Including figure files
\usepackage{amsmath}	% Advanced maths commands
\usepackage{amssymb}	% Extra maths symbols
\usepackage{xspace} 
\usepackage{xcolor}
\usepackage[caption = false]{subfig}
\usepackage{ulem}
\newcommand{\msun}{\,$\mathrm{M}_\odot$\xspace}%
\newcommand{\gcm}{\,g~cm$^{-3}$\xspace}%
\newcommand{\astrod}{\texttt{astrodendro}\xspace} %
\newcommand{\yt}{\texttt{yt}\xspace} %
\newcommand{\unsim}{\mathord{\sim}}
\newcommand{\orion}{{\sc Orion2}\xspace}

\newcommand{\changed}[1]{{{}#1}}

\title[The time evolution of cores]{The highly variable time evolution of star-forming cores identified with dendrograms}

\author[Smullen et al.]{
Rachel~A.~Smullen$^{1},$\thanks{E-mail: rsmullen@email.arizona.edu}
Kaitlin~M.~Kratter$^{1}$, 
Stella S. R. Offner$^{2}$, 
Aaron T. Lee$^{3}$, 
\newauthor
Hope How-Huan Chen$^{2}$
\\
$^{1}$Steward Observatory, University of Arizona, Tucson, AZ 85721, USA\\
$^{2}$Department of Astronomy, University of Texas Austin, Austin, TX, 78712, USA\\
$^{3}$Department of Physics and Astronomy, Saint Mary's College of California, Moraga, CA 94575, USA}
\date{Accepted XXX. Received YYY; in original form ZZZ}

\pubyear{2019}

\begin{document}

\label{firstpage}
\pagerange{\pageref{firstpage}--\pageref{lastpage}}
\maketitle

\begin{abstract}
We investigate the time evolution of dense cores identified in molecular cloud simulations using dendrograms, which are a common tool to identify hierarchical structure in simulations and observations of star formation.  We develop an algorithm to link dendrogram structures through time using the three-dimensional density field from magnetohydrodynamical simulations, thus creating histories for all dense cores in the domain. We find that the population-wide  \textit{distributions} of core properties are relatively invariant in time, and quantities like the core mass function match with observations. Despite this consistency,  an \textit{individual} core may undergo large (>40\%), stochastic variations  due to the redefinition of the dendrogram structure between timesteps.  This variation occurs independent of  environment and stellar content.  We identify a population of short-lived (<200 kyr) overdensities masquerading as dense cores that may comprise $\unsim20\%$ of any time snapshot.  Finally, we note the importance of considering the full history of  cores when interpreting the origin of the initial mass function; we find that, especially for systems containing multiple stars, the core mass defined by a dendrogram leaf in a snapshot is typically less than the final system stellar mass. This work reinforces that there is no time-stable density contour that defines a star-forming core.  The dendrogram itself can induce significant structure variation between timesteps due to small changes in the density field. Thus, one must use caution when comparing dendrograms of regions with different ages or environment properties because  differences in dendrogram structure may not come solely from the physical evolution of dense cores.
\end{abstract}


%%%%%%%%%%%%%%%%%%%%%%%%%%%%%%%%%%%%%%%%%%%%%%%%%%

%%%%%%%%%%%%%%%%% BODY OF PAPER %%%%%%%%%%%%%%%%%%

\section{Introduction}
Understanding the early progression and end state of star formation is fundamental to many areas of astronomy, from modeling the formation of galaxies to studying the assembly of planetary systems. Stars form in dense molecular cores embedded within gravitationally contracting filamentary structures  \citep{Andre2010,Arzoumanian2013,Smith2014,Arzoumanian2019}. On core scales, gravity sets the dominant dense core properties \citep{Goodman2009,Lee2014,Storm2016}, while turbulence \changed{is thought to regulate} the star formation efficiency and core formation, including properties like core rotation \citep{Padoan2012,Chen2018b}.  There is also a population of observed, pressure-confined cores that will likely not form stars if left untouched, although these objects may later collapse due to shock interactions \citep{Seo2015,Keown2017,Kirk2017,Chen2019}. 
Despite an understanding of this broad process of star formation, there are still many open questions. These include the relationship between observed core masses and the initial mass function, the time evolution of dense core properties, the role of the physical environment in the star formation process, and the formation mechanisms of bound binary (or higher order multiple) systems, among others.  

Previous works have attempted to answer some of these questions by looking at individual snapshots of observed regions or simulations, yet few have ever attempted to correlate the evolution of individual cores with the broad core property distributions reported in the literature. The interplay between the time evolution of individual cores and their contribution to distributions of core properties may be especially important when understanding the connection between the core mass function (CMF) and stellar initial mass function (IMF) \citep[][and references therein]{Offner2014}.  There is still debate about whether the IMF directly inherits its shape from the CMF 
%This set of citations is causing the PDF to break if not wrapped in an mbox
%\mbox{\citep[e.g.,][]{Padoan2002,Hennebelle2008,Hopkins2013}} 
\citep[e.g.,][]{Padoan2002,Hennebelle2008,Hopkins2013}
or is independent of core masses \citep[e.g.,][]{Bonnell2001a,Bate2003,Clark2007}. The IMF is frequently fit with the form of a power law at high masses and a log-normal distribution at lower masses as first demonstrated in \cite{Chabrier2003}.  Subsequent work has suggested that the IMF is mostly independent of star-formation physics such as accretion rate and star formation inefficiency  \citep{Hennebelle2012,Cunningham2018}, but may depend on local environmental properties like the global radiation field and local magnetic fields \citep{Offner2009,Bate2009,Dib2017,Guszejnov2017,Lee2017,Cunningham2018,Ntormousi2019}.  Thus, it is imperative to know how individual cores may contribute to the interplay between the CMF and IMF evolution.

A fundamental aspect to properly interpreting both simulation snapshots and observations of star-forming regions is understanding \textit{what} overdensities are identified \changed{as cores}. Core identification in both observed star-forming regions and simulations has been a topic of active investigation for decades.  Beginning with the by-eye identification of structure in molecular clouds from \cite{Blitz1986}, the field has expanded in two dominant directions.  The first direction is the singular identification of dense clumps, which started from the watershed segmentation algorithm of \cite{Williams1994}.  This developed into the \textsc{clumpfind} algorithm that has been utilized extensively. The other method of core identification is using hierarchical structure methods. \changed{Other core-finding methods that return singular clumps include the gradient-tracing scheme {FellWalker}  \citep{Berry2015}, {GaussClumps} \citep{Stutzki1990}, which fits Gaussians to all peaks in the data, and  \textsc{cutex} \citep{Molinari2011}, which looks for curvature changes in the data,  among others.}  Early hierarchical structure methods such as the structure trees from \cite{Houlahan1992} then evolved into the commonly adopted dendrogram algorithm first presented in \cite{Rosolowsky2008}. Dendrograms connect structures in star-forming regions from filaments to dense cores and allow a better understanding of the hierarchical nature of the star formation process \citep{Goodman2009}. 

Each core identification algorithm comes with its own often subtle biases that must be understood in the context of the analysis performed \citep[e.g.,][]{Li2019}. \changed{For instance, GaussClumps and \textsc{cutex} only fit elliptical sources, but GaussClumps can easily handle overlapping sources \citep{Stutzki1990} and \textsc{cutex} works well with large background variations \citep{Molinari2011}.  \textsc{clumpfind} has been found to be sensitive to input parameters but is widely available \citep[e.g.,][]{Berry2015}. FellWalker clumps can sometimes have artificial splitting due to the cleaning process but tends to be more robust to noise \citep{Berry2015}. Dendrograms can be sensitive to the algorithm tuning choices but provide the best understanding of the physical environment surrounding cores \citep{Rosolowsky2008}.  The above is not a comprehensive list of the benefits and drawbacks of core identification methods, but it serves to show that every algorithm in use will work better in some situations as compared to others. }

Simulations have become a critical tool to interpret the necessarily-incomplete window provided by observations in star-forming regions, especially as simulations have grown in resolution and complexity. For instance, \cite{Mairs2014} note the importance of high resolution observations in recovering the full mass and detailed structure of star-forming cores.  Observations at moderate resolution tend to miss mass and structure due to averaging errors  \citep{Offner2012,Mairs2014}. Similarly, \cite{Beaumont2013} reach the important conclusion that position-position-velocity observations carry uncertainties of 40\% in computed quantities when compared to a three-dimensional simulation.
Effects like gas superposition \changed{along the line of sight}, line opacity \changed{obscuring core structure}, and mapping observational \changed{(PPV)} space to physical \changed{(PPP)} space contribute confusion to an accurate \changed{physical} interpretation of cores from observations, \changed{because the line-of-sight structure of a core can be easily miscalculated} \citep[e.g.,][]{Ostriker2001,Ballesteros-Paredes2002,Shetty2010,Beaumont2013}.

In this paper, we explore the time evolution of star-forming cores identified with dendrograms and work to understand the role of the dendrogram algorithm itself in the properties of identified cores. We begin by creating an algorithm to link dendrogram structures through time, which we describe in Section~\ref{s:methods}. The robustness of this methodology depends on several tunable parameters, and we explore the effect of the three major parameters in Section~\ref{s:testing}. We present the time evolution of the identified cores, the distributions of core properties, and other results of note in Section~\ref{s:results}, and we then explore the reasons for the variability we find in  Section~\ref{s:dendros}. Finally, Section~\ref{s:discussion} notes  the implications of our findings, including the importance of full core histories and the limitations of the dendrogram algorithm. Section~\ref{s:conclusions} summarizes our findings.

\section{Methodology}\label{s:methods}

In this work, we aim to trace the histories of cores in simulations of star formation to test the robustness of core parameters measured throughout a core's lifetime. Here, we discuss the magnetohydrodynamic simulation used in this work, describe our method of core identification, and present our new core tracking method.    \changed{We describe the fiducial choices for our core identification and linking methodology in this section and test the impact of variations in these choices in Section~\ref{s:testing}.}

%\vspace{0.1in}
\changed{\subsection{A note on nomenclature}

The meaning of the word ``core'' is not well defined between works. Observationally-based definitions of cores, adapted from \cite{Chen2019}, include ``dense cores'', which are regions that have a dominant thermal velocity and low virial $\alpha$ (virialized), ``starless cores'' that do not have a protostar and are not virialized, ``prestellar cores'' that do not have a protostar but are virialized, and ``protostellar cores'', which do have protostars and are virialized.  Cores can also be defined in simulations as the material that accretes onto a star particle \citep[e.g.,][]{Bate2003}, or the region of dense material at a single snapshot \citep[e.g.,][]{Ntormousi2019a}.
 
This paper explores differences between algorithmic, physical, and phenomenological understandings of dense cores in star formation.  Toward that goal, we use the term \emph{leaf} for a dense structure inside contours identified with a dendrogram, the term \emph{overdensity} for a physical collection of dense gas in the simulations, and the term \emph{core} for the loosely-defined, observationally motivated dense structures that may form stars.

}
\subsection{Simulations}\label{ss:simulations}

Our simulation initial conditions are identical to those of run W2T2 in \citet{OffnerArce2015}. These conditions are intended to model a piece of a local, Gould Belt star-forming region like the Perseus molecular cloud. For our purposes, the simulation represents a prototypical turbulent molecular cloud that serves as a test-bed for our core identification and tracking method; the properties of the cloud itself have little bearing on our methodology as we are investigating trends in the evolution of core structure. We outline the initialization and parameters of the simulation below.

We run a magnetohydrodynamical (MHD) simulation of a $\unsim3800$\msun ($7.5\times10^{36}\textrm{\,g}$) gas cloud using the \orion code \citep{Klein1999,Krumholz2007a,Li2012}.  \orion is a 3-dimensional adaptive mesh refinement (AMR) MHD grid code that includes physics such as self-gravity, ideal MHD \citep{Li2012}, and Lagrangian accreting sink particles \citep{Krumholz2004,Lee2014a}. Our simulations are initialized on a $256^3$ base grid that corresponds to 5\,pc on a side with periodic boundary conditions in all spatial dimensions. \changed{We expect little influence on core evolution from our choice of boundary condition as compared to a global molecular cloud simulation.}

These simulations refine the spatial resolution based on the Jeans number $J$ such that 
\begin{equation}\label{eq:jean}
J\equiv\frac{\Delta x\changed{_i}}{\lambda_J}<0.125\text{,}
\end{equation}
where $\Delta x_i$ is the cell size at the current level $i$ and $\lambda_J~=~(\pi c_s^2/G\rho)^{1/2}$ is the Jeans length \citep{Truelove1997}.  When $J>0.125$, finer cells with size $\Delta x_i$ are added, thus resolving the Jeans wavelength with higher resolution.  Our simulations have 5 refinement levels over the base grid, which defines our minimum resolution as $\unsim4034 \textrm{ AU}$ per cell for the $256^3$ grid at level 0 and the maximum resolution as $\unsim126 \textrm{ AU}$ per cell for the cells refined to level 5.  These sizes are defined based on the mean gas density in the simulation of $\rho_0=2\times10^{-21}$\gcm and the mean sound speed of $18800\,\textrm{cm\,s}^\changed{-1}$.  In regions undergoing gravitational collapse, gas is removed from the grid and replaced with a sink particle if $J>0.25$ on the finest level \citep{Krumholz2004}. Sinks accrete mass and momentum from gas within a radius of four cells at level 5 as well as interact gravitationally with the surrounding gas.

We generate the cloud initial conditions through a turbulent driving phase that proceeds without gravity, which produces self-consistent turbulent gas density and velocity distributions \citep[e.g.,][]{maclow1999,li2004,Offner2008}. 
The simulation begins with a uniform density, uniform temperature of 10~K and a uniform magnetic field in the $z$ direction, $B_z= 13.5\,\mu$G, which corresponds to an initial thermal pressure to magnetic pressure ratio (plasma $\beta$) of $\beta = 8 \pi \rho_0 c_s^2/B_z^2$ = 0.1.
Then the gas is stirred for two gas crossing times by perturbing the gas velocities with a random velocity distribution that corresponds to a flat distribution in Fourier space with wave numbers $k=1-2$. At the end of the driving phase, the gas reaches a turbulent steady state with a turbulent power spectrum $P(k)\propto k^{-2}$ and $\beta=0.02$ \citep{OffnerArce2015,Offner2018}.
Finally, self-gravity is turned on, and we evolve the simulation for approximately 70\% of a global free-fall time ($t_{\rm ff} = \sqrt{3 \pi/ 32 G \rho_0} \simeq$ 1.5 Myr). 

We adopt a barotropic equation of state of the form $p = \rho c_{\rm iso}^2 [1.0 + (\rho/\rho_{\rm c})^{\gamma-1}]$, where $c_{\rm iso}$ is the sound speed for 10~K gas, $\rho_{\rm c}$ is the critical density at which the gas transitions from isothermal to adiabatic and $\gamma=5/3$ is the adiabatic index.  We choose an effective critical density that is comparable to the Jeans density on the maximum AMR level, $\rho_{\rm c}=7 \times 10^{-15}$\,g cm$^{-3}$. This value is smaller than the expected critical density for dense gas, $\rho_{\rm c} \simeq  10^{-14}$\,g cm$^{-3}$ \citep{masunaga1998}, in order to produce some warming when the maximum gas densities are reached. This lower critical density acts to eliminate contiguous small scale fragmentation, which would otherwise occur within isothermal filaments at high resolution \citep{Kratter2010} and roughly approximates the influence of radiative feedback, which is expected to heat the gas once protostars form \citep{Offner2009}. 

\begin{figure}
\centering
\includegraphics[width=\columnwidth]{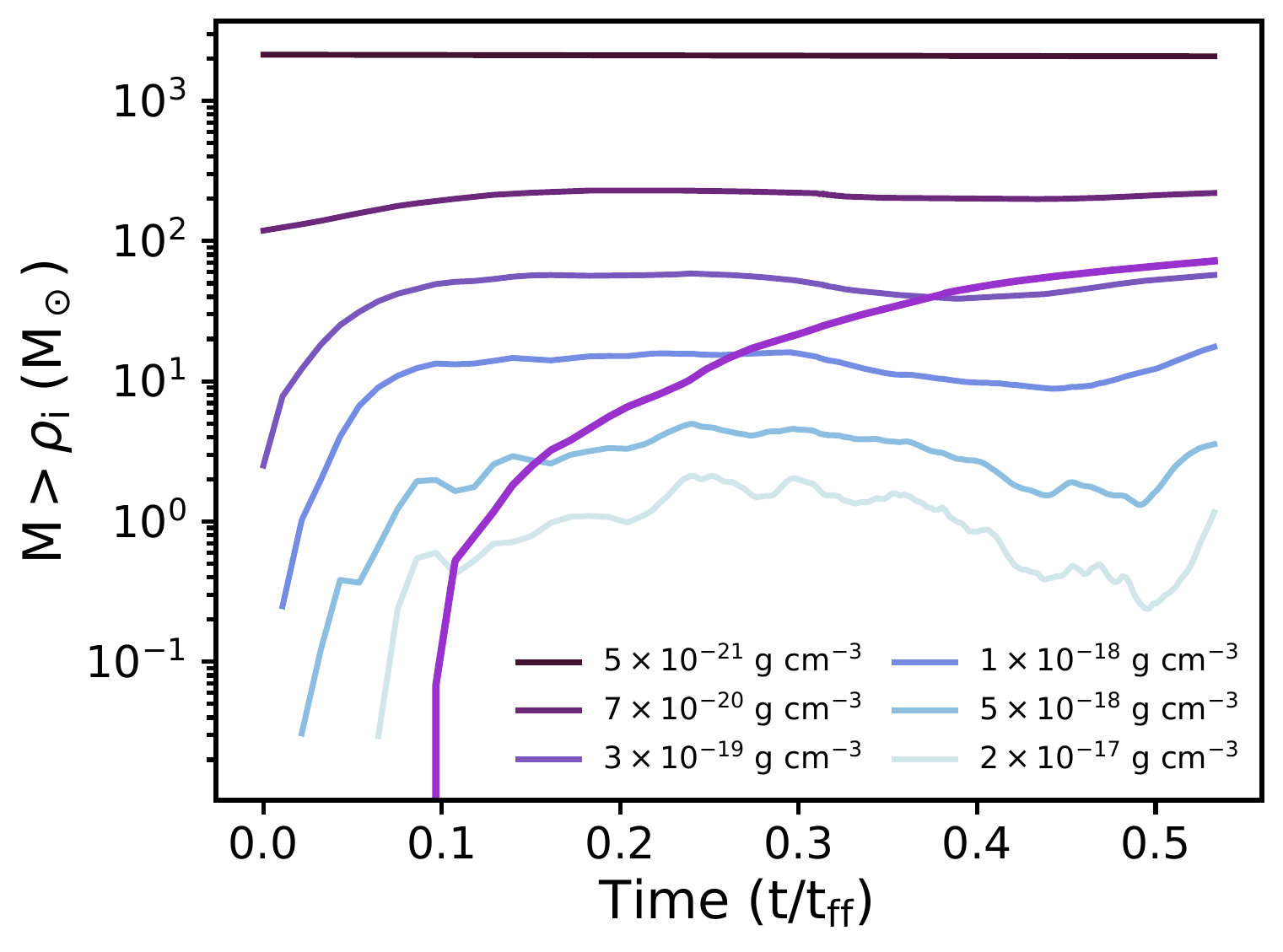} % plot_mass_density_thresh.py 
\caption{Time evolution of dense gas in our simulation. The blue colored lines indicate the total mass above a given density across time, while the purple line shows the mass of sink particles across time. \changed{The lowest density shown, $5\times10^{-21}$\gcm, corresponds to a number density of approximately 1300 nH$_2 \textrm{ cm}^{-3}$, while the highest density of $2\times10^{-17}$\gcm corresponds to a molecular hydrogen density of $5\times10^6 \textrm{ cm}^{-3}$.  }
\label{f:mvt}}
\end{figure}

The time evolution of the dense gas is shown in Figure~\ref{f:mvt}.  The density thresholds used in this plot correspond to the densities of the AMR refinement thresholds (computed from Equation~\ref{eq:jean}).  The simulations end with about 70\msun of mass in 25 sink particles.  The fraction of mass in the densest gas shows more variability because of mass accretion onto the sink particles.  \changed{Most of the dense cores are formed in one large filament that spans the majority of the volume.}

\subsection{Structure identification}\label{ss:dendro}

Dendrograms are a common tool used to identify dense structures in both simulations and observations of star-forming regions; many previous works have used them to find and identify properties of bound clumps and filaments \citep[e.g.,][]{Rosolowsky2008,Goodman2009, Beaumont2013,Burkhart2013,Lee2014,Seo2015,Storm2016,Friesen2016,Keown2017,Wong2017,Nayak2018,Chen2018}. Dendrograms also provide a metric to quantify the structure of molecular cloud emission and associated physical properties \citep[e.g.,][]{Boyden2016,Koch2017,Boyden2018,Koch2019}.  \changed{For example, \citet{Goodman2009} demonstrated that dendrograms produce more physically reasonable identifications of cores in 3D spectral line data compared to another previously popular algorithm, {\sc clumpfind}.  As an additional benefit, dendrograms naturally identify nested features and therefore reflect the relationship between structures of different sizes in the data.}
Thus, we choose dendrograms as our structure identification algorithm to better quantify the interpretation of this widely adopted method. \changed{However, because of the fundamental similarities between all core identification methodologies--that cores are identified from peaks in quantities such as density or emission--the results of this work should be generally applicable across algorithms.}

A dendrogram is a tree \changed{algorithm} that identifies hierarchical structures in any input quantity in a 2- or 3-dimensional grid.  A dendrogram contains leaves (the most refined structure), trunks (the lowest level structure that may contain refined substructure), and branches (structures that connect leaves to other branches or trunks). The dendrogram initialization is commonly defined by three parameters: the background level cutoff that defines the base of the tree, the minimum difference (height) between two nested structures in the quantity being dendrogrammed required to create a new branch or leaf, and the minimum size of an identified structure.  The dendrogram is built by first identifying the maximum value in the grid.  The algorithm then iteratively searches adjoining cells and uses the size and density increase criteria to determine if a new branch or leaf needs to be created.  The tree ends when the background cutoff value is reached.  Neighboring leaves can be children of a single branch if they are both contained in the spatial bounds of the branch.  A region can contain multiple \changed{unconnected} trees if an area surrounding two structures is below the background cutoff. Note that dendrograms are inherently relative structures because they are computed based on the maximum value in a region. This work utilizes the \astrod Python package.\footnote{http://www.dendrograms.org/}

We carefully consider how we optimize the three parameters (background cutoff, minimum density increase for new structure creation, and minimum structure size) that define how a dendrogram is built\footnote{In the \astrod package, these variables are named \texttt{min\_value}, \texttt{min\_delta}, and \texttt{min\_npix}}.  The background cutoff will set the fraction of gas in  the simulation included in the dendrogram and impact the total height of the tree (peak to minimum density, which will likely also impact the number of structures in a tree) and the number of branches that can be created.  The minimum density increase to create new structure sets a height at which new branches and leaves are created: a smaller value allows smaller density increases to be considered as new structure, while a larger value makes the creation of new structure much more \changed{stringent}.  Finally, the minimum structure size influences the size and internal complexity of an identified core.  Too large of a size means that we might group individual compact structures into one leaf, while too small of a size might over-resolve substructure  in the star-forming cores we wish to study (e.g., lumps in a disk-like overdensity).

\changed{The background density threshold is a major limitation to the complexity of the dendrogram. A low background density threshold connects more of the cloud structure, including filaments, but these structures would not be readily observable in traditional tracers. On the other hand, a high density threshold might prematurely truncate low density structure.}
We use a fiducial density threshold of $\rho=7\times10^{-20}$\gcm, which is the density needed to refine a cell from level 0 to level 1. The threshold density chosen herein roughly corresponds to the minimum density observed in ammonia emission ($n\gtrsim10^4 \textrm{\,cm}^{-3}$ or $\rho=4\times10^{-20}$\gcm), so the structures identified in our dendrograms would be observed in synthetic observations \citep{Flower2006}. This creates a dendrogram that contains only a few percent of the data by volume and consistently contains the same dense structures throughout the entire simulation time.  As seen in Figure~\ref{f:mvt}, our fiducial density encompasses a roughly constant mass (around $100$\msun) over the length of the simulation.  \changed{Variations of the background density cutoff are described in Section~\ref{ss:cutoff}.}

We next consider the density increase to create a new leaf.  This parameter impacts the inclusion of low-density structures in our dendrograms. \changed{A small density increment produces many nested structures}, and these new  structures (typically intermediate branches) do not add to the understanding of either \changed{leaf} structure or the hierarchy.  \changed{A large density increment leads to very large leaves and begins} to under-resolve the dense structures \changed{that best resemble dense cores by combining multiple overdensities into one leaf}.  We therefore chose a factor of 3 increase in density as the fiducial density contrast required to create a new leaf.  This \changed{choice} is further discussed in Section~\ref{ss:contrast}.

Finally, we set the fiducial minimum size of a structure to be 125 voxels (3-dimensional cells; at our fiducial grid size, one voxel is $(1000\,\textrm{AU})^3$). \changed{The minimum size of 125 voxels} is large enough to encompass compact structures, such as protostellar disks, without being  \changed{small enough to allow clumpy sub-structures to split} into multiple leaves.
%{\bf KMK: previous sentence doesn't make sense to me}  
At our fiducial resolution, this leads to a \emph{minimum} leaf size of about $(5000~\textrm{\,AU})^3$, although most structures are substantially larger.   For comparison, the dendrograms of \cite{Friesen2016}, who investigate the size and mass of embedded clumps in the Serpens South protocluster, have a smallest effective radius of $\unsim0.02\textrm{\,pc}$ or $\unsim 4100 \textrm{\,AU}$.  Typical observed core sizes from works such as \cite{Seo2015} and \cite{Keown2017} are $10^{-2}\textup{--}10^{-1} \textrm{\,pc}$, so our minimum size resolves structures in our simulation that are similar to observed cores.  The choice of the other two \changed{dendrogram initialization} parameters \changed{can, in some instances}, negate the utility of the minimum size.  If the background density threshold is high and the density increase is large, small structures will be not be able to be resolved and every structure will exceed the minimum structure size.  With our fiducial parameters, structure can approach the minimum size but the majority of leaves have volumes of a few hundred to a few thousand voxels \changed{(core sizes $\gtrsim0.05 \textrm{\,pc}$)}.

While dendrograms can be computed for any scalar quantity, we choose to compute dendrograms on the \changed{three-dimensional density grid}. The large dynamic range of physical density in our simulations means that a logarithmic scaling better traces the physical structures. Therefore, to define structures in our simulations, we compute a dendrogram with periodic (wrapping) boundary conditions on the log of the density field at each simulation snapshot.  The dendrogram routine can only search a uniform grid, so we must apply a covering grid to our AMR simulations.  Covering grids interpolate the AMR data onto a fixed grid of size $256\cdot2^{i}$ in each dimension where 256 is the base size of the simulation and $i$ is the level for which we want to create the grid. We define our dendrogram on a level 2 covering grid with each cell having a side length of $1.5\times10^{16}$ cm, or about $0.005 \textrm{pc}\approx 1000\textrm{ AU}$. Our choice of fiducial resolution is discussed in Section~\ref{ss:resolution}. 

\begin{figure*}
\centering
\includegraphics[width=\textwidth]{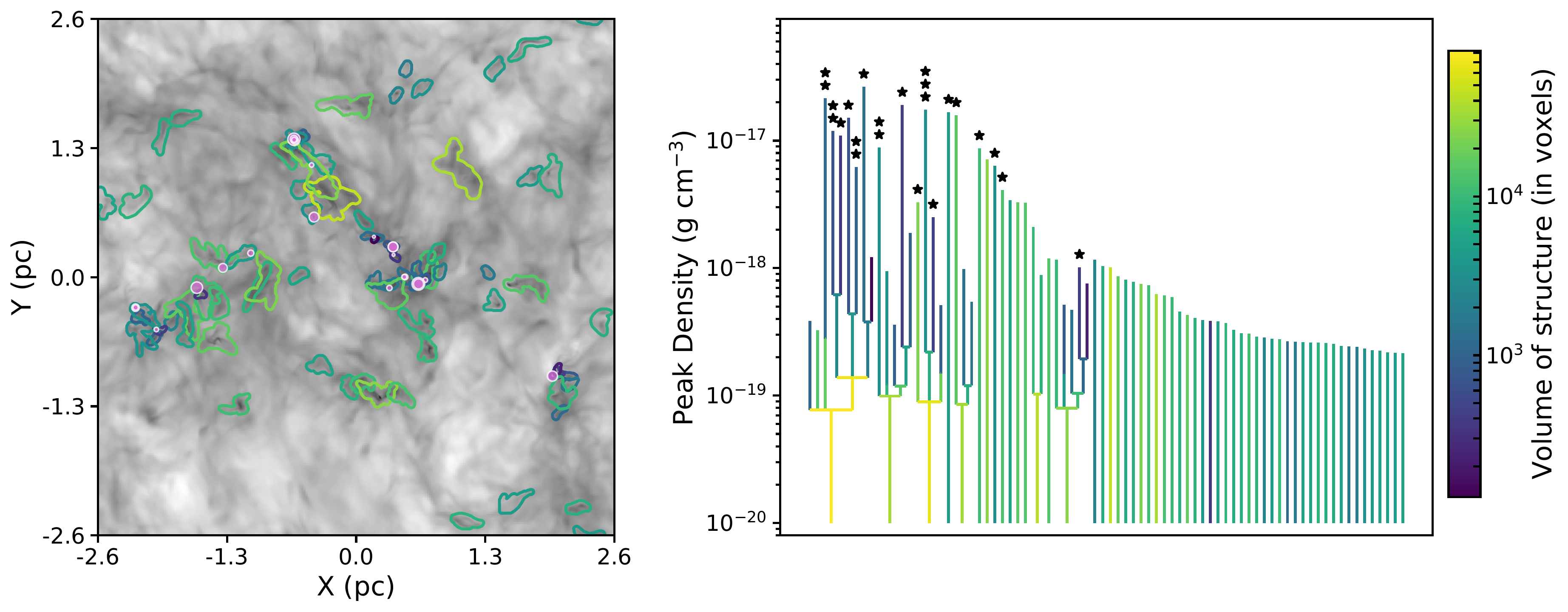} %  plot_dendrocontours_new.py
\caption{  An example of the dendrogram computed with fiducial parameters at an intermediate timestep.  The right panel shows the tree colored by the volume of the leaf.  Black stars denote the presence of sink particles in the leaf.  The $x$-axis has no physical meaning; the structures have been roughly sorted by peak density.  The left panel shows the $x-y$ projection of the leaves. Pink circles denote the location of sink particles 
\changed{with sizes scaled by the mass of the sink}.
\label{f:tree}}
\end{figure*}

Our choice of parameters leads to dendrograms containing about 80 leaves in all but the earliest timesteps when gravitating structure has barely started to collapse.  An example of the dendrogram computed with these fiducial parameters at an intermediate timestep (40\% $t_\textrm{ff}$) is shown in Figure~\ref{f:tree}.   Many of the leaves \changed{($>50\%$)} are isolated and not part of a larger structure that contains further refinement due to our choice of background density cutoff; however, these leaves tend to be of relatively low-density, and most will likely not form stars as they are temporary structures (see further discussion in Section~\ref{ss:noise}). Most leaves \changed{($>80\%$)} do not form sink particles by the end of the simulation, and leaves containing one sink particle are about 2-3 times more common than leaves hosting multiple sink particles. Sink-hosting leaves \changed{can decrease in peak density} after sink formation due to accretion of high-density gas onto the sink particle. 

With the dendrogram defined, we output a catalog of important leaf parameters at each timestep (density, position, velocity, magnetic fields, etc.) using the full AMR grid that falls within the volume of the uniform-grid leaf surface.  This catalog is then used to perform the linking algorithm defined in the next section.

\subsection{Linking structures through time}\label{ss:linking}

Once the 3-D structures are constructed for every timestep, we link them through time. 
We take a two step approach by first linking structures between consecutive timesteps and then by reconstructing a structure's  full path through time.  

\subsubsection{Pair-wise linking}\label{sss:pairwise}

To match structures between timesteps, we use a geometric search that relies upon simulation outputs being frequent enough that structures do not move significantly (more than about half of their size) between outputs.  Beginning with an individual leaf ($l_{a}$) at timestep $t_a$, we search for all leaves at timestep $t_b$ where the center of mass of $l_{a}$ is within the surface of a leaf at $t_b$.  We then reverse the search such that we look for the center of mass of a time $t_b$ leaf to be within the surface of a time $t_a$ leaf.  We do allow for an offset of the center of mass from the boundary of the leaf in two dimensions because of the possibility of dendrogram contours being defined differently between consecutive timesteps as discussed below.  The choice of this offset is described in Section~\ref{ss:link_dist}, but our fiducial value is set to 10 grid cells at level 2.  A leaf at one timestep can be associated with multiple leaves at the next timestep, and we describe the consequences of this further below. This search is then continued between all consecutive pairs of timesteps (i.e., $t_a \leftrightarrow t_b$, $t_b \leftrightarrow t_c$, etc.).

\begin{figure}
\centering
\includegraphics[width=0.5\columnwidth]{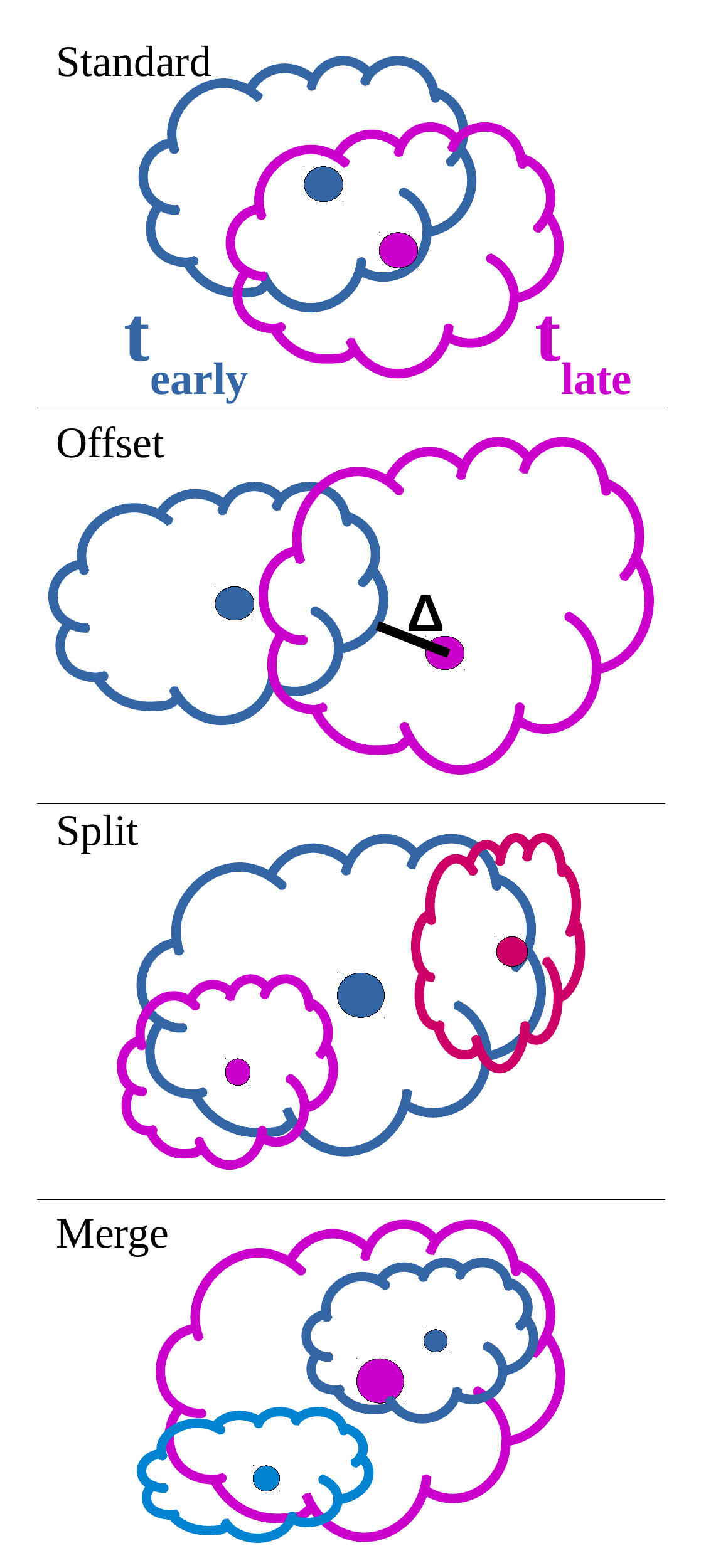}
\caption{ A cartoon of the cases that result from the pair-wise linking. Blue colors indicate earlier times and the purple colors indicate later times.  The top is the ``standard'' linking where the center of mass at one timestep (filled circle) is found within the volume at the other timestep (open contour).  The ``offset'' linking allows there to be a small offset ($\Delta$) between the leaf center of mass and the leaf volume at consecutive timesteps, which typically arises from dendrogram contours being redrawn to include more material. ``Splits'' and ``mergers'' are cases in which a leaf at one timestep can be associated with more than one leaf at a consecutive time.  Note that, while this cartoon is shown in 2-D, the linking in our data is done in 3-D.
\label{f:cartoon}}
\end{figure}

There are \changed{four} cases that result from the pair-wise linking as shown in Figure~\ref{f:cartoon}.  Leaves can be uniquely identified with a single structure between $t_a\rightarrow t_b$ and $t_b\rightarrow t_a$. This is most common and leads to a single path between timesteps (panels ``standard'' and ``offset'').  However, multiple  leaves can be found at one timestep that map back to a single leaf at the adjacent timestep.  If the single leaf is at an earlier timestep and the multiple leaves are at a later timestep, this is a ``split''.  If the single leaf is at a later timestep and the multiple leaves are at an earlier timestep, this is a ``merger''.  In our simulations, splits and mergers are most frequently due to dendrogram leaf boundaries being drawn to  encompass multiple nearby overdensities, not actual physical merging or fragmentation. \changed{Physical evolution can happen but is difficult to disentangle from the changes in dendrogram contours.}

\changed{In the 170 output timesteps of our simulation, we link 11,000 leaf pairs; 10,500 of those are securely linked, meaning that we identify the same linked pair looking forward and backward.  We find $\unsim200$ splits and mergers. About 60--70\% of linked pairs have no offset, and the average offset of the remaining linked pairs is about 10 cells ($\unsim0.05$ pc). }

\begin{figure}
\centering
\includegraphics[width=1\columnwidth]{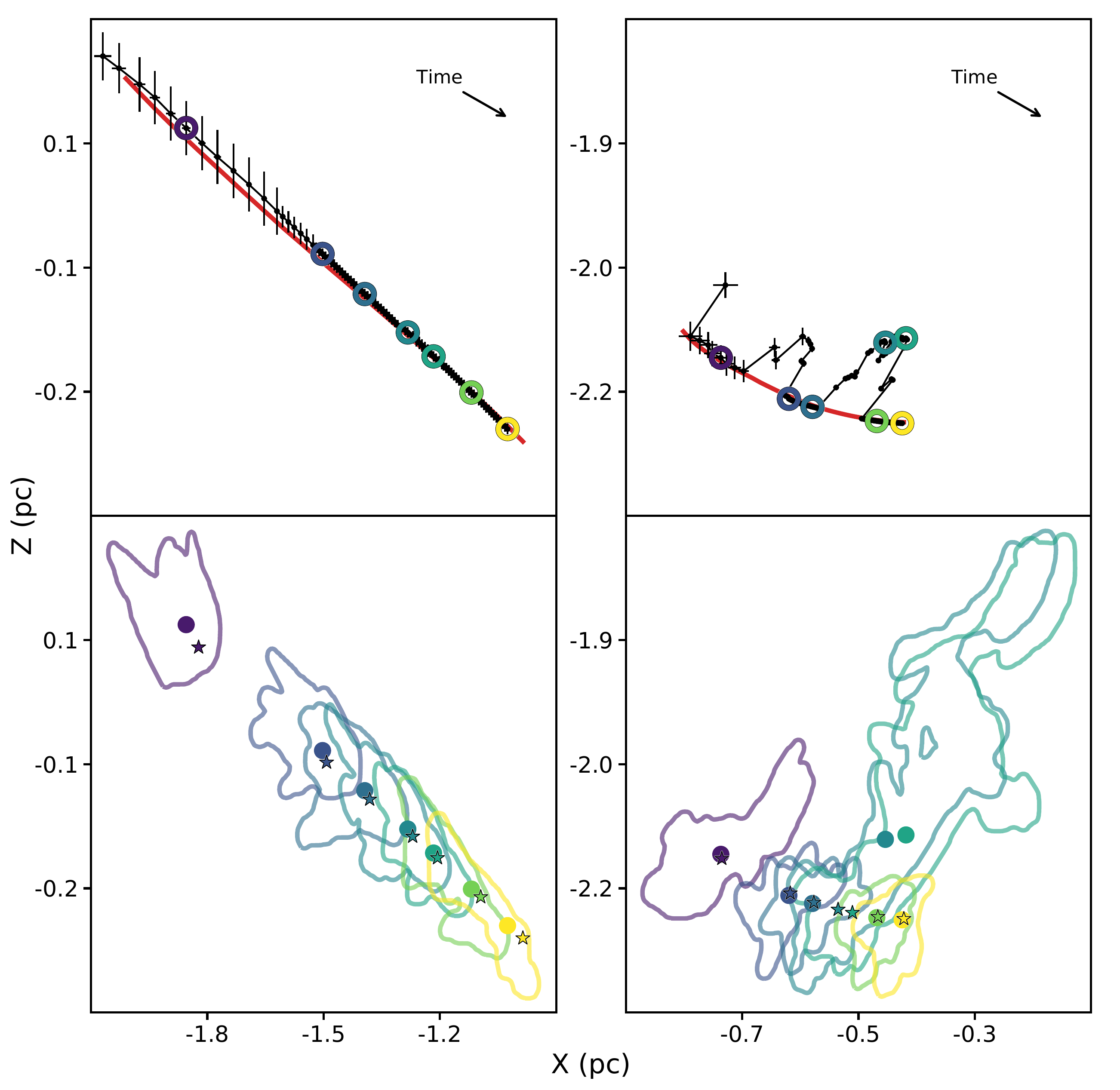} % find_leaves_for_sink1.py 
\caption{ Two examples of leaf behavior illustrated by sink-hosting leaves.  The left column shows an ideal behavior where the leaf structure evolves smoothly, while the right column shows a leaf that undergoes significant dendrogram structure variations in time. The upper panels show the $x-z$ positions of a leaf through time.  The black line show the center of mass of the leaf through time, while the red line shows the position of the sink particle in the leaf.  The error bars on the black line show the ``expected'' position of the leaf at the next timestep given the velocity of the center of mass of the leaf. The bottom panels show a selection of the projected leaf contours at the times indicated by colored points in the top panel. The solid circles denote the center of mass of the leaf and the star indicates the position of the sink particle.  The lower right panel demonstrates a common barrier to velocity-based linking: the dendrogram leaf contours can change significantly between timesteps, offsetting the center of mass of the leaf from the expected position.  
\label{f:tracking}}
\end{figure}

We initially incorporated the velocity information, specifically the leaf center-of-mass velocity, \changed{(which includes the contribution from any sink particles that may be present in the leaf)}, into our algorithm to help inform the direction of motion to uniquely track a leaf through time and reduce the number of nearby, unassociated leaves linked.  However, because of variability in the computation of dendrogram structures \changed{between timesteps}, the leaf center of mass does not always move in predictable patterns. Therefore the addition of velocity information does not improve our linking.  To demonstrate this issue, we present two examples of leaf behavior in Figure~\ref{f:tracking}.  The error bars \changed{in the upper panels have been doubled in length to be more visible.}   The  left panels shows what a well-behaved leaf looks like: the leaf center of mass at the next time is within the position expected from the velocity.  However, a significant number \changed{($\gtrsim 25\%$)} of our \changed{leaves have a history that} look more like the right panels, where at certain times, the dendrogram contours are redrawn to include more material. This then changes the center of mass of the leaf and the expected position of the leaf center is wildly offset from the computed location of the center of mass.  The leaf centers at consecutive timesteps are typically within the leaf contours, meaning that the less complicated geometric search discussed above is more reliable for our data.  We do encounter \changed{a few} pathological cases where a reconstruction cannot be performed in an automated way, such as when the leaf is shaped like a banana-- the center of mass lies outside the leaf contour and is therefore computed to have a large offset to the leaf boundary at the next timestep.

\subsubsection{Path reconstruction}\label{sss:reconstruction}

The last step to fully trace the histories of overdensities in our simulations is to transform the pairwise linking into a coherent path through time.  We use the terminology ``path'' to denote a single set of related leaves through time and ``path family'' to denote a group of paths that were found to be related to a single starting leaf. For the analysis presented herein, we work backward in time (from the end of the simulation to the start) because the most relevant structures to compare to observations are the compact overdensities found in well-evolved regions \changed{at later times in the integration}. \changed{Because we use a fixed starting point in time, cores may be traced at different evolutionary stages.}

 We start by selecting a leaf ($l_{0}$) from the cohort of leaves at the final timestep ($t_0$). We search through the linked pairs $t_0 \leftrightarrow t_1$ to find the leaf at $t_1$ linked to $l_{0}$.  This found leaf is added to the path. We then check if  the leaf at $t_1$ is associated with any other leaves at $t_0$.   We then iteratively repeat this process to search for matches to the earliest leaf in the path going backward in time.

For the cases where there are mergers (two or more leaves at an earlier time being associated with only one leaf at a later time), we add one of the leaves to the current path and then add new paths to the path family by copying the current path and appending the other merged leaf.  Each path in the path family is then reconstructed independently.

For the cases where there are splits (two or more leaves at a later time being associated with only one leaf at an earlier time), we create a new path and recursively search in the opposite direction (from early times to late times) to find the path(s) associated with the new leaf.

Path families can have many component paths because each new split or merger effectively doubles the number of paths in a path family.  While not always indicative of physical interactions, a large-number path family does indicate that the structure lives in a crowded area of the simulation volume.

\section{Parameter Variations}\label{s:testing}

\changed{All core identification algorithms include tunable parameters, and the dendrogram algorithm we adopt here is no exception. In observational studies, the parameters are chosen based on the noise, sensitivity and resolution of the data. When analyzing simulation data there is more flexibility in parameter choice. Consequently, we explore a variety of parameter values to assess the physical impact of our parameter choices, including background cutoff density,  density increment to create a new structure, grid resolution, and linking distance $\Delta$.}  Thus, in this section, we explore how variations in these parameters impact our reconstructed path families.  While the minimum leaf volume at constant resolution is also a tunable parameter, we find negligible impact on the final dendrogram structure when varying this quantity within reasonable limits.

\changed{
\subsection{Background cutoff}\label{ss:cutoff}
The background cutoff influences the tree complexity, leaf structure, and computational requirements of a dendrogram. Values near the mean density in the simulation ($5\times10^{-21}$\gcm) include too much gas that never participates in the star formation process.  Very large, low density structures affiliated with the filamentary structure are commonly identified as leaves.  Values at high-density (level 2 refinement density or higher, or around $3\times10^{-19}$\gcm) exclude an extremely large portion of the gas ($\gtrsim99\%$), including gas at early times that will eventually fall in to a dense core.  The level 2 refinement density corresponds to a Jeans length of 0.08 pc, which is smaller than typically observed cores.  The dendrogram would be less likely to resolve any structure larger than this, which includes most of the objects that resemble observed cores; instead, the algorithm would only identify small peaks in larger overdensities. For these reasons, and the physical arguments described in the previous section, we use the fiducial density of $7\times10^{-20}$\gcm.
}

\subsection{Density increment}\label{ss:contrast}

\begin{figure*}
\centering
\includegraphics[width=0.8\textwidth]{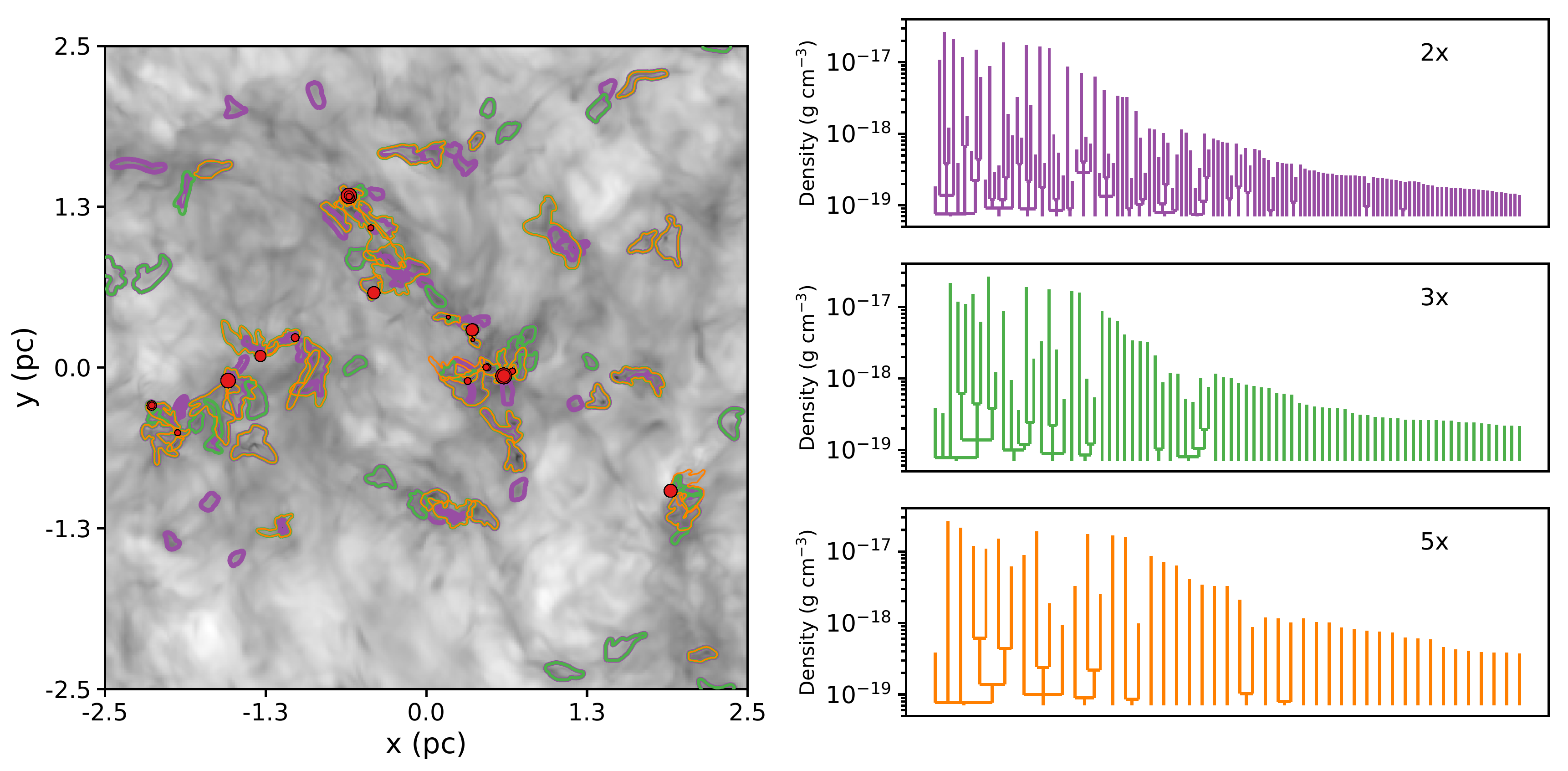} %plot_comparison_2x3x5x_withsidedendros.py
\caption{ A comparison of dendrogram leaves given different density increments.  On the left, the purple, green, and orange contours demarcate leaves from dendrograms with a 2x, 3x, and 5x density increase (contrast) required to create a new leaf, respectively.  Contours are shown over a density projection of the simulation with darker colors indicating denser regions.  Red circles show the locations of sink particles and are scaled by sink mass.  The right panels show the full dendrograms for each density increment.  The different trees trace out the same dense material, indicating that our results are relatively invariant of the choice of contrast.  
\label{f:contrast}}
\end{figure*}

The contrast required to create a new structure in the dendrogram mainly impacts the low-density structure identified in the tree. We compare density increases of factors of 2, 3 \changed{(our fiducial choice)}, 4, 5, and 10 and find little difference between the leaves identified, although the trees themselves are quite different.  We show the comparison of factors of 2, 3, and 5 in Figure~\ref{f:contrast}.  All high-density structure is contained in all trees; the \changed{major} differences arise in the low-density structures.  Every sink particle lives in a nearly identical leaf, meaning that the important structures for star formation are not impacted \changed{greatly} by our choices of density contrast parameter. 

\changed{The trees computed with large density contrasts identify much less structure because overdensities must be much more significant to be added to the tree.  This means that, in dense regions especially, two neighboring overdensitites may be enclosed in one leaf.  Small density contrasts can lead to very low density leaves being added to the tree.  These leaves are insignificant temporary perturbations above the background cutoff and add a level of unnecessary ``noise'' to the linking process.}

\cite{Burkhart2013} perform a similar analysis by varying the density increase required to create a new structure ($\delta$ in that work) and comparing the resulting dendrograms across a suite of MHD simulations.  They find that dendrogram structure varies significantly with $\delta$ and can provide information about the relative importance of shocks, self-gravity, and super-Alfv\'{e}nic turbulence.

\subsection{Resolution}\label{ss:resolution}

The size and shape of structures in the dendrogram are highly correlated with the resolution of the uniform grid used to compute the dendrogram.  When all other parameters  are kept fixed, an increase in resolution, unsurprisingly, allows for both more refined structures and physically smaller structures.  The algorithm identifies more structures because each increase in the level of the uniform grid provides a factor of 8 increase in the number of cells, meaning that there is more flexibility to define compact structure.  Physically smaller structures are identified because the minimum size of a structure is fixed at 125 cells; therefore, each increase in level decreases the minimum required physical volume of a structure by a factor of 8. 

We compute our fiducial dendrogram at level 2. $1024^3$ cells in the volume gives a $1000 \textrm{AU/cell}$ resolution.  We also tested level 3 ($2048^3$ cells; $500 \textrm{AU/cell}$) and level 4 ($4096^3$ cells; $250 \textrm{AU/cell}$) resolution.  The memory required for producing a dendrogram at the  full level 4 volume was prohibitive.  We therefore  use a subset of the volume of size $\left[2048,2048,1536\right]$ at level 4 (about 10\% of the volume) that contains 15 of our 24 sink particles and many of the structures identified at level 2. Other than resolution, we compute the dendrograms for each case using the same fiducial parameters as described above in Section~\ref{ss:dendro}.  The comparison volume contains 40 level 2 leaves, 52 level 3 leaves, and 63 level 4 leaves.

\begin{figure*}
\centering
\includegraphics[width=0.8\textwidth]{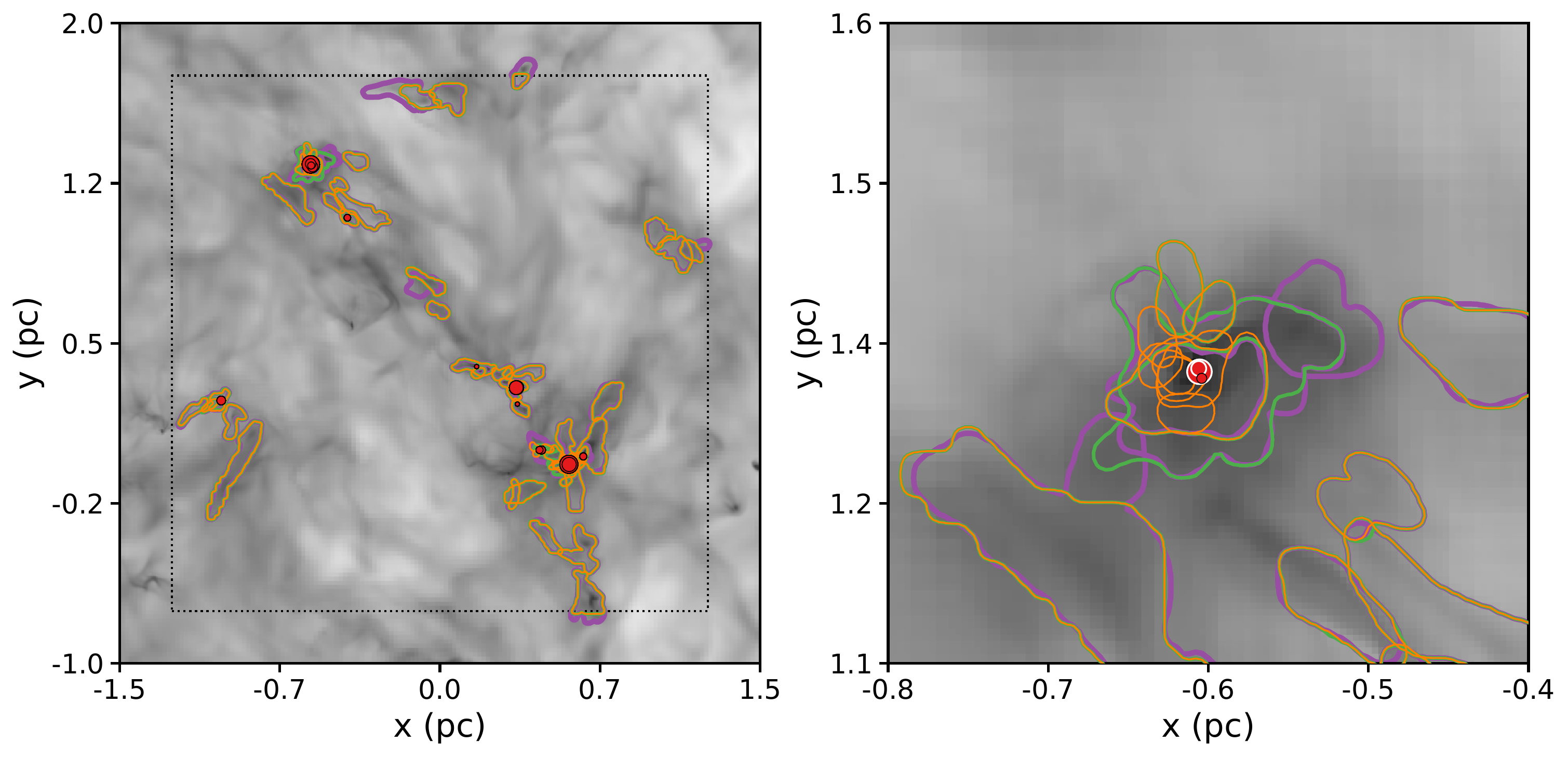} % density_3E-19_smallbox/plot_comparison_l2l3l4.py
\caption{ Contours of leaves from trees computed on uniform grids at AMR levels 2 (the fiducial choice in this work), 3, and 4 in purple, green, and orange, respectively.  The contours are shown over a density projection of the simulation with darker colors indicating denser regions.  Red circles show the locations of sink particles \changed{and are scaled in size relative to their mass}.  The black dashed line in the left panel indicates the extent of the comparison volume.  As is seen in the left panel, the contours at all levels broadly agree.  The right panel reveals that only in the densest regions do the leaf volumes differ significantly. \changed{This region contains an overdensity that surrounds a bound triple system.  The sink particles outlined in white comprise the central binary (5\msun and 2\msun), while the small circle outlined in black is the tertiary companion (0.8\msun).  The tertiary is separated from the central binary by a few thousand AU.}
\label{f:level_234}}
\end{figure*}

The contours of leaves at the three levels are shown in Figure~\ref{f:level_234}.  As seen in the left panel, the contours at all levels broadly agree.  Only in the densest regions do the leaf volumes differ significantly.  The right panel shows one of these dense regions: a triple system in a complicated overdensity illustrates how differences in the resolution can change the leaf structure.  The level 2 dendrogram encloses all triple members in one leaf.  The level 3 dendrogram draws the central binary in one contour but excludes the tertiary component along the z-direction.  The tertiary's local overdensity is not large enough to create an independent leaf.  The level 4 dendrogram assigns all the small overdensities in the greater disk-like overdensity to their own leaves. 

Figure~\ref{f:level_234} suggests that level 4 is too sensitive to substructure:  the overdensity of one physically bound system is \changed{often split into sub-structures} such that we lose information about the \changed{bound core}.  It is not possible to compute important system quantities such as gravitational potential or virial parameter without having the full bound structure contained in a single leaf.  

\begin{figure}
\centering
\includegraphics[width=\columnwidth]{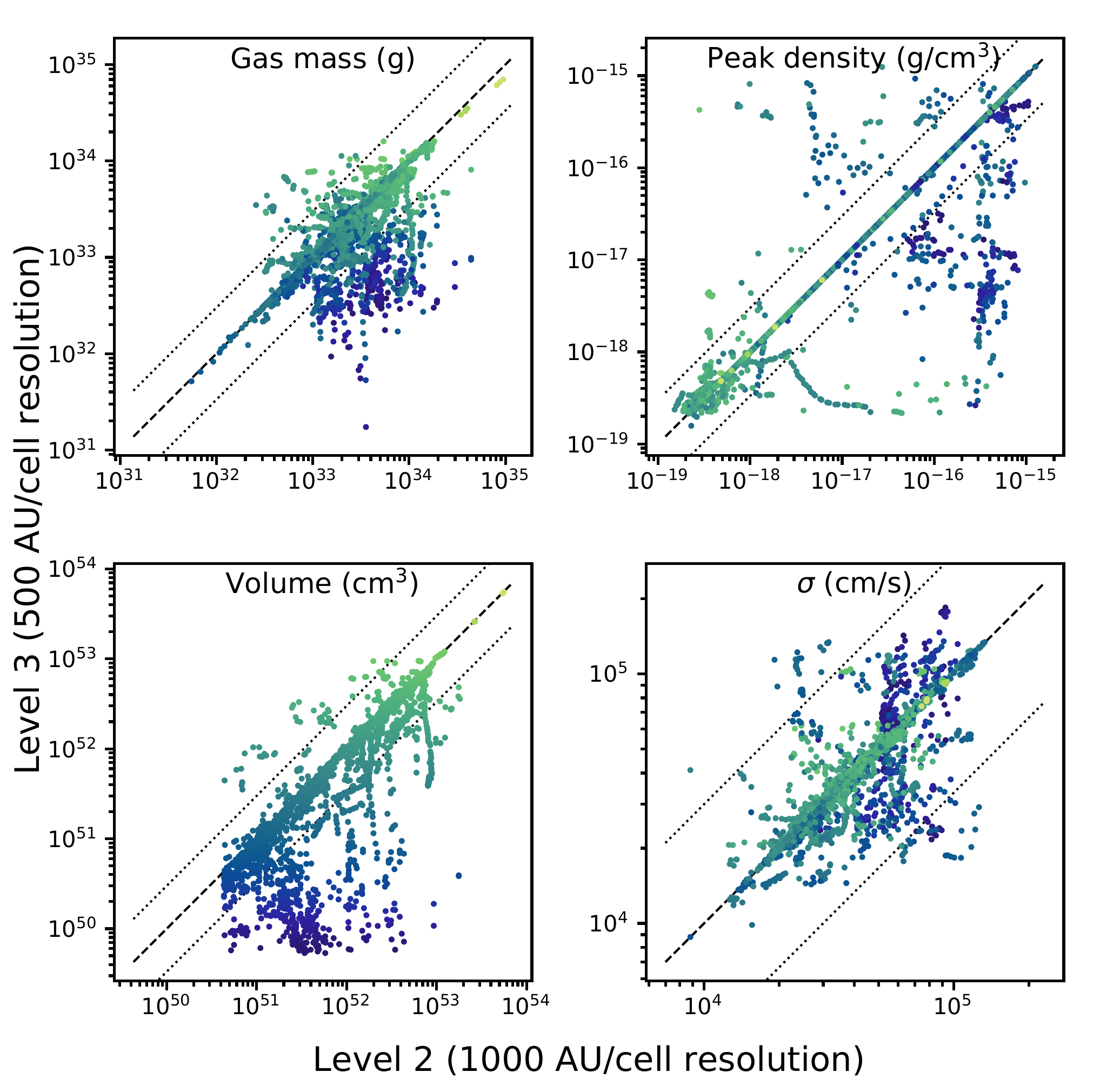} %density_3E-19_smallbox/level3_3x/compare_level_quants.py
\caption{ A comparison of derived leaf parameters between correlated leaves at level 2 and level 3.  The panels show gas mass, peak density, volume, and velocity dispersion. Points are colored by the volume at level 3. The horizontal axis shows the values of leaves computed on level 2 and the vertical axis shows the value for the corresponding leaf computed at level 3. The black dashed line shows the one-to-one correlation, while the dotted lines show a factor of three difference.  The correlation in mass, peak density, and velocity dispersion is very good despite the fact that there is a factor of 8 difference in the minimum leaf volume.
\label{f:quants_l2l3}}
\end{figure}

Next, we asses the utility of the level 3 dendrogram. Ideally, we want to minimize large changes in the density contours while including all important structure (or sink particles) in a given overdensity. To this end, we compare the derived leaf parameters  between leaves tracing the same physical structure at level 2 and level 3 in Figure~\ref{f:quants_l2l3}. The correlation in mass, peak density, and velocity dispersion is very good (typically within a factor of 3) despite the fact that there is a factor of 8 difference in the minimum volume.  \changed{Outliers below the one-to-one correlation in all panels except velocity dispersion arise from the population of small, low-density leaves that are identified as independent structures at level 3.}  These small overdensities are typically included as part of a larger level 2 leaf at the lower resolution, but the low density of the structures means that the mass-weighted parameters such as velocity dispersion are mostly agnostic about their inclusion. \changed{Leaves above the one-to-one line in the same panels come from ambiguities in the correlation of leaves across resolution, but they \changed{constitute} a small fraction of the total number of leaves shown. Patterns of points forming lines in any of the panels are indicative of time evolution.} Thus, we conclude that the difference in leaves identified at level 2 and level 3 does not impact our understanding of dense core evolution. 

\subsection{Linking distance}\label{ss:link_dist}
\changed{While the aforementioned parameters  control the construction of the dendrogram at each time snapshot, the linking distance is the crucial parameter that controls the history of structures ($\Delta$ in the  ``offset'' panel in Figure~\ref{f:cartoon}).} The linking distance is the distance between the center of mass of a leaf at one timestep and the surface of the leaf to which it has been linked.  Linking distance will simultaneously impact the number of paths in a family and the number of timesteps traced in an individual path.  Typical leaf sizes are of order 0.2 pc, so we test linking distances of 0 cells (no offset; a leaf center is within the contour at the neighboring time), 10 cells ($\unsim 10^4\textrm{ AU}$; about a \changed{typical} leaf radius), 100 cells ($\unsim 10^5\textrm{ AU}$; about 10 leaf radii), and 200 cells ($\unsim 2\times 10^5\textrm{ AU}$).  Our goal is to robustly identify leaves with common histories without permitting too many uncertain connections while, at the same time, allowing for variations in dendrogram leaf contours.  

\begin{figure}
\centering
\includegraphics[width=\columnwidth]{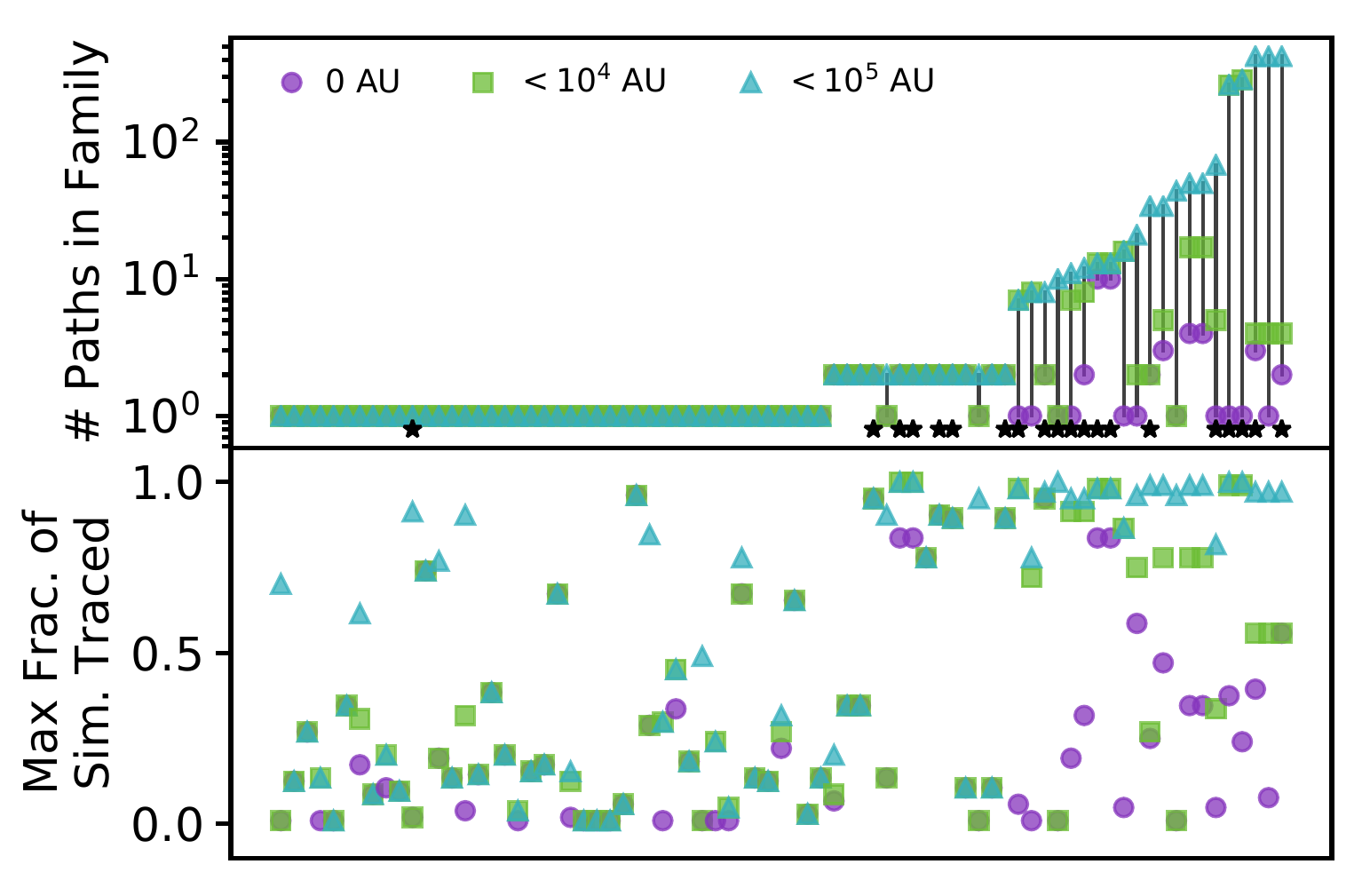}% link_test/compare_reconstructs.py
\caption{ Comparison of different linking distances.  The colored shapes represent different linking distances.  The top panel shows the number of paths in a family (the unique combinations of leaf histories identified for a single starting leaf).  The horizontal axis is arbitrary and simply serves to order the leaves. Vertical lines connect the the path families for a single leaf at different linking distances.  Black stars along the bottom indicate the presence of at least one sink particle in a leaf. The bottom panel shows the maximum fraction of the total simulation time traced by a path family; ordering matches the top panel. The choice of linking distance is most significant in dense environments where there are many leaves in close proximity.
\label{f:reconstruct}}
\end{figure}

We present the results of our investigation in Figure~\ref{f:reconstruct}.  There are minimal variations between the 100 cell and 200 cell linking distances, so we only present the 100 cell results in the figure.   For most path families, specifically those of isolated leaves, linking distance does not make a difference in the number of paths reconstructed. The smaller two linking distances, on average, create smaller numbers of paths in the family.  Some of this is due to large variations in dendrogram contours between timesteps; the changes in the leaf boundaries can be larger than the linking distance. Linking distance does have a much stronger impact on the length of a path history, however. Larger linking distances typically lead to longer paths, while a linking distance of 0 can sometimes artificially truncate a path.

The linking distance becomes an important consideration for leaves in dense environments, which is also where the majority of sink particles reside, including the many bound multiple systems.  In these environments, leaves can exhibit significant variation in structure between timesteps, and therefore a very small linking distance will result in frequent premature truncation of paths.  However, because of the proximity to many dense structures, it is easy to link two nearby, but not physically interacting, structures, leading to a large increase in the number of paths in a family.

We adopt 10 grid cells as our fiducial linking distance because it allows some variation in the dendrogram leaf contours without leading to linking with many nearby, unassociated leaves.  We are still able to identify paths through a substantial fraction of the simulation time, but we don't reach the extremely numerous, \changed{and less physically meaningful,} path families found with larger linking distances.

\section{Results: Core properties and evolution}\label{s:results}

We have identified dense core \changed{analogs} using dendrograms at multiple simulation outputs and reconstructed the time evolution of these leaves. We now study the broad distributions of leaf properties (such as is frequently done in other work, both observational and computational) and the time evolution of individual leaves in our simulations.  We summarize our findings below.

\subsection{Core property distributions vs.~individual core history}\label{ss:dist}

\begin{figure}
\centering
\includegraphics[width=\columnwidth]{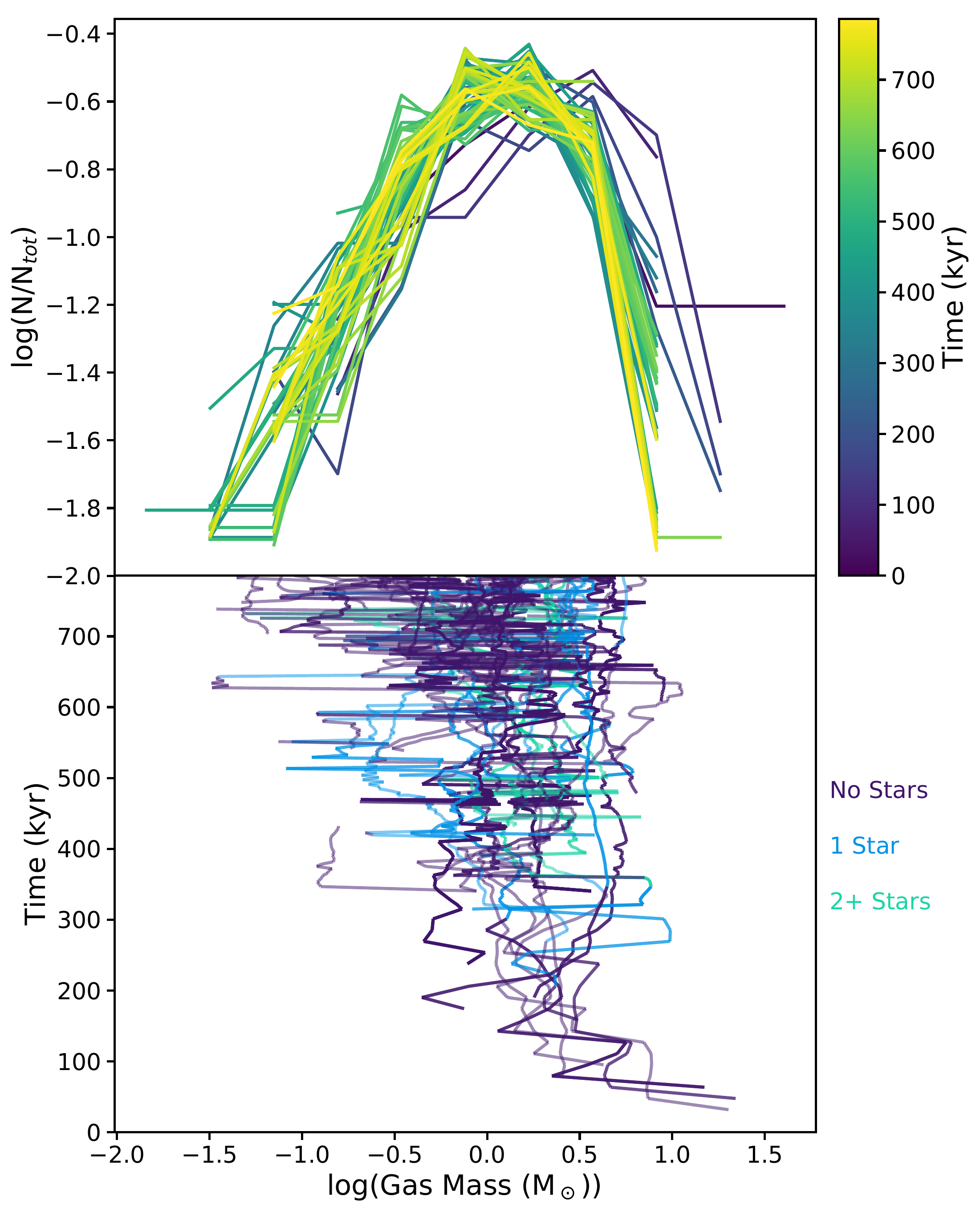} %plot_cmf_withlines_andtime.py
\caption{ Core mass distribution vs. individual core evolution.  The top panel shows the distribution of gas mass in the cores across time.  The bottom panel shows the gas mass evolution of a subset of reconstructed paths through time; dark purple lines show leaves without sinks particles, blue lines show leaves with a single sink particle, and green lines show leaves with multiple sink particles.  While the broad distribution of gas mass is nearly invariant in time, any individual leaf may have large variations in its reconstructed  history. 
\label{f:dist}}
\end{figure}

We study both the distribution and individual evolution of the large sample of leaves in our simulation. Every parameter distribution we investigate is relatively constant in time.  However, the individual evolution of a leaf can be quite variable. We present an example of this dichotomy in Figure~\ref{f:dist}.  The mass distribution does not vary significantly in time; major variations are only seen at the earliest times when structure is beginning to collapse due to self-gravity.   The bottom panel reveals that a leaf may have a computed gas mass that can span upwards of an order of magnitude \changed{in time}, and the typical evolution does not smoothly vary from time to time. Note that in these (and all future figures), time is measured relative to the beginning of the simulation.

To better understand the relative variability in core evolution, we use a parameter called the coefficient of variation (CV), which is defined as the standard deviation of a parameter ($\sigma$) divided by the mean ($\mu$) of that parameter.  This quantity allows us to directly compare leaf properties of varied units and physical scales and has units of percent.  For our analysis, we consider paths that are tracked for more than 15 kyr and compute the standard deviation of the total path.  Because of the rapid evolution in both volume and density of leaves at early times due to the introduction of gravitationally collapsing structure, we exclude the earliest $\unsim 30\%$ of the simulation from our computation of the CV.

Table~\ref{tab:CVs} presents the minimum, mean, and maximum variation of 16 different parameters: total mass, gas mass, leaf volume, leaf size, mean density, oblateness, virial parameter, the Mach number of gas in the core, the Mach number of the core in the simulation volume, the Alfven Mach number of the core, the angular momentum magnitude, the variation in the angular momentum orientation, the magnetic field magnitude, the variation in the magnetic field orientation, and plasma $\beta$.  The definitions of these parameters are presented in Appendix~\ref{definitions}. \changed{While there are few substantial trends to remark upon for individual quantities, for completeness, we report CVs for the entire ensemble of parameters studied in our analysis.} We have separated the paths into three bins in each section.  Under each bin is the number of individual paths that fall into the bin.  The final line in each table section is $\Delta$, \changed{ which is the spread in CV (maximum CV minus minimum CV) for the collection of quantities and is designed to show the variation in variability for each bin.}

\subsubsection{Size of path family}\label{sss:size}
We first split our full contingent of paths by the number of paths in a family.  Paths with $n=1$ are isolated; they typically show the least variation.  However, these paths are, on average, shorter than paths in other bins that can lead to suppressed variation. Paths with $n\ge10$ are typically in very dense environments and are therefore most susceptible to being linked to multiple nearby leaves.  This can cause variations to be artificially high as physically unassociated cores \changed{(overdensities that don't physically interact in space)} are linked in the same path; the large CV of volume in the non-isolated paths hints that leaf contour changes (arising from structures bouncing above and below the structure refinement threshold due to minute changes in the local density field)    
may cause the large CVs in other parameters.

\subsubsection{Number of sinks} \label{sss:number}
We then group paths by the number of sinks in the leaf at the final timestep.  Paths with $n=0$ are starless overdensities. The starless paths with low CVs are typically short-lived, low-density leaves. The paths containing multiple sink particles are frequently part of large path families in dense regions where physically independent overdensities are identified as related due to a temporary co-location or dendrogram contour changes that cause multiply-linked leaves. 

\subsubsection{Length of path}\label{sss:length}
Finally, we divide the ensemble of paths by their length.  The shortest paths \changed{($<75\textrm{kyr}$)} are frequently temporary overdensities and therefore have little physical evolution over the time they are traced as indicated by the low CVs. However, there are also short paths with high variability that belong to a large path family. The longest paths \changed{(lifetimes greater than 250kyr)} have little correlation with the size of the path family or the presence of stars, so the CVs in the final bin span a large range. 

\subsubsection{General trends}\label{sss:general}
The path histories we trace have significant variation-- frequently upwards of 40\% in CV. In all three methods of dividing paths presented in the table, the average CV increases from left to right, meaning that shorter lived, isolated, starless cores tend to have less variability.  However, the maximum CV does not show the same trend, indicating that any given path can vary significantly.

It is also important to note that a low variability in one parameter does not indicate low variability in all parameters. This is demonstrated with the parameter $\Delta$ at the bottom of each section.  This quantity shows the maximum difference in the CV of the 16 parameters for each leaf, or the maximum variation in the variation of our computed properties.  The average $\Delta$ in all cases is over 50\%, meaning that the majority of paths have little correlation in the amount of variability in different quantities. Thus, the computed properties of observed overdensities identified with dendrograms may not correlate well with the physical evolution of the bound core itself.

{
\tabcolsep=0.11cm
\begin{table*}
\centering
\caption{Coefficient of variance range for different core properties.  The CV is defined as the standard deviation divided by the mean of a quantity. \label{tab:CVs}}
\resizebox*{!}{\textheight}{%
\begin{tabular}{l|ccc|ccc|ccc} 
\multicolumn{10}{c}{Number of paths in a family}\\ 
\hline 
   &  \multicolumn{3}{c}{$n=1$}  &  \multicolumn{3}{c}{$1<n<10$}  &  \multicolumn{3}{c}{$n\geq10$}\\ 
   &  \multicolumn{3}{c}{(34)}  &  \multicolumn{3}{c}{(89)}  &  \multicolumn{3}{c}{(790)}\\ 
\hline 
Quantity  &  Min  &  Mean  &  Max  &  Min  &  Mean  &  Max  &  Min  &  Mean  &  Max\\ 
  &  (\%)  &  (\%)  &  (\%)  &  (\%)  &  (\%)  &  (\%)  &  (\%)  &  (\%)  &  (\%)\\ 
\hline 
Total mass  &  9  &  \textbf{27}  &  53  &  12  &  \textbf{55}  &  213  &  9  &  \textbf{42}  &  153\\ 
Gas mass  &  9  &  \textbf{27}  &  53  &  10  &  \textbf{57}  &  279  &  13  &  \textbf{52}  &  144\\ 
Volume  &  1  &  \textbf{17}  &  61  &  10  &  \textbf{68}  &  266  &  18  &  \textbf{67}  &  152\\ 
Size  &  1  &  \textbf{31}  &  277  &  4  &  \textbf{52}  &  246  &  6  &  \textbf{27}  &  243\\ 
Mean density  &  2  &  \textbf{17}  &  96  &  3  &  \textbf{56}  &  196  &  12  &  \textbf{67}  &  244\\ 
%Pressure  &  2  &  \textbf{17}  &  99  &  3  &  \textbf{57}  &  199  &  10  &  \textbf{68}  &  277\\ 
Oblateness  &  1  &  \textbf{26}  &  89  &  16  &  \textbf{43}  &  92  &  15  &  \textbf{38}  &  74\\ 
Virial $\alpha$  &  3  &  \textbf{27}  &  75  &  8  &  \textbf{49}  &  214  &  14  &  \textbf{41}  &  254\\ 
Internal $\mathcal{M}$  &  1  &  \textbf{10}  &  34  &  4  &  \textbf{24}  &  65  &  4  &  \textbf{22}  &  49\\ 
Total $\mathcal{M}$  &  0  &  \textbf{6}  &  20  &  2  &  \textbf{11}  &  40  &  5  &  \textbf{13}  &  28\\ 
Alfven $\mathcal{M}$  &  2  &  \textbf{11}  &  30  &  6  &  \textbf{32}  &  104  &  11  &  \textbf{30}  &  72\\ 
$\left|\vec{j} \right|$  &  3  &  \textbf{38}  &  154  &  18  &  \textbf{77}  &  242  &  10  &  \textbf{80}  &  189\\ 
(Max($\vec{j}$)-Min($\vec{j}$))/Mean($\vec{j}$)  &  0  &  \textbf{25}  &  64  &  5  &  \textbf{27}  &  53  &  11  &  \textbf{31}  &  61\\ 
$\left|\vec{B} \right|$  &  1  &  \textbf{12}  &  56  &  1  &  \textbf{27}  &  54  &  4  &  \textbf{24}  &  60\\ 
(Max($\vec{B}$)-Min($\vec{B}$))/Mean($\vec{B}$)  &  1  &  \textbf{17}  &  45  &  6  &  \textbf{29}  &  71  &  10  &  \textbf{31}  &  48\\ 
Plasma $\beta$  &  2  &  \textbf{16}  &  67  &  6  &  \textbf{51}  &  156  &  14  &  \textbf{43}  &  140\\ 
\hline 
$\Delta$  &  19  &  \textbf{59}  &  272  &  31  &  \textbf{110}  &  270  &  36  &  \textbf{94}  &  262\\ 
\hline 
\multicolumn{10}{c}{Number of sinks at final timestep}\\ 
\hline 
   &  \multicolumn{3}{c}{$n=0$}  &  \multicolumn{3}{c}{$n=1$}  &  \multicolumn{3}{c}{$n>1$}\\ 
   &  \multicolumn{3}{c}{(520)}  &  \multicolumn{3}{c}{(85)}  &  \multicolumn{3}{c}{(308)}\\ 
\hline 
Quantity  &  Min  &  Mean  &  Max  &  Min  &  Mean  &  Max  &  Min  &  Mean  &  Max\\ 
  &  (\%)  &  (\%)  &  (\%)  &  (\%)  &  (\%)  &  (\%)  &  (\%)  &  (\%)  &  (\%)\\ 
\hline 
Total mass  &  9  &  \textbf{54}  &  213  &  9  &  \textbf{44}  &  213  &  21  &  \textbf{25}  &  110\\ 
Gas mass  &  9  &  \textbf{54}  &  279  &  11  &  \textbf{51}  &  207  &  24  &  \textbf{48}  &  157\\ 
Volume  &  1  &  \textbf{61}  &  266  &  7  &  \textbf{53}  &  190  &  28  &  \textbf{77}  &  196\\ 
Size  &  1  &  \textbf{39}  &  277  &  6  &  \textbf{29}  &  246  &  13  &  \textbf{14}  &  31\\ 
Mean density  &  2  &  \textbf{53}  &  185  &  17  &  \textbf{61}  &  196  &  18  &  \textbf{83}  &  244\\ 
%Pressure  &  2  &  \textbf{52}  &  188  &  18  &  \textbf{62}  &  199  &  21  &  \textbf{88}  &  277\\ 
Oblateness  &  1  &  \textbf{36}  &  89  &  29  &  \textbf{51}  &  92  &  15  &  \textbf{38}  &  59\\ 
Virial $\alpha$  &  3  &  \textbf{49}  &  141  &  9  &  \textbf{44}  &  96  &  15  &  \textbf{28}  &  254\\ 
Internal $\mathcal{M}$  &  1  &  \textbf{26}  &  55  &  5  &  \textbf{19}  &  61  &  4  &  \textbf{15}  &  65\\ 
Total $\mathcal{M}$  &  0  &  \textbf{14}  &  25  &  3  &  \textbf{11}  &  28  &  6  &  \textbf{11}  &  40\\ 
Alfven $\mathcal{M}$  &  2  &  \textbf{19}  &  104  &  8  &  \textbf{28}  &  72  &  15  &  \textbf{47}  &  57\\ 
$\left|\vec{j} \right|$  &  3  &  \textbf{64}  &  189  &  10  &  \textbf{56}  &  242  &  21  &  \textbf{109}  &  185\\ 
(Max($\vec{j}$)-Min($\vec{j}$))/Mean($\vec{j}$)  &  0  &  \textbf{28}  &  64  &  8  &  \textbf{30}  &  61  &  22  &  \textbf{35}  &  53\\ 
$\left|\vec{B} \right|$  &  1  &  \textbf{20}  &  56  &  11  &  \textbf{22}  &  43  &  8  &  \textbf{33}  &  60\\ 
(Max($\vec{B}$)-Min($\vec{B}$))/Mean($\vec{B}$)  &  1  &  \textbf{26}  &  71  &  12  &  \textbf{32}  &  50  &  13  &  \textbf{37}  &  45\\ 
Plasma $\beta$  &  2  &  \textbf{37}  &  79  &  3  &  \textbf{38}  &  154  &  14  &  \textbf{53}  &  156\\ 
\hline 
$\Delta$  &  19  &  \textbf{90}  &  272  &  45  &  \textbf{85}  &  241  &  53  &  \textbf{105}  &  262\\ 
\hline 
\multicolumn{10}{c}{Length of path history}\\ 
\hline 
   &  \multicolumn{3}{c}{$t<75 \textrm{kyr}$}  &  \multicolumn{3}{c}{$75 \textrm{kyr}<t<250 \textrm{kyr}$}  &  \multicolumn{3}{c}{$t>250 \textrm{kyr}$}\\ 
   &  \multicolumn{3}{c}{(35)}  &  \multicolumn{3}{c}{(21)}  &  \multicolumn{3}{c}{(857)}\\ 
\hline 
Quantity  &  Min  &  Mean  &  Max  &  Min  &  Mean  &  Max  &  Min  &  Mean  &  Max\\ 
  &  (\%)  &  (\%)  &  (\%)  &  (\%)  &  (\%)  &  (\%)  &  (\%)  &  (\%)  &  (\%)\\ 
\hline 
Total mass  &  9  &  \textbf{54}  &  213  &  10  &  \textbf{42}  &  81  &  9  &  \textbf{43}  &  153\\ 
Gas mass  &  9  &  \textbf{62}  &  279  &  10  &  \textbf{41}  &  81  &  11  &  \textbf{51}  &  144\\ 
Volume  &  1  &  \textbf{60}  &  266  &  2  &  \textbf{36}  &  102  &  5  &  \textbf{67}  &  152\\ 
Size  &  1  &  \textbf{29}  &  217  &  6  &  \textbf{44}  &  171  &  6  &  \textbf{29}  &  277\\ 
Mean density  &  2  &  \textbf{35}  &  196  &  3  &  \textbf{37}  &  110  &  5  &  \textbf{65}  &  244\\ 
%Pressure  &  2  &  \textbf{36}  &  199  &  3  &  \textbf{37}  &  110  &  4  &  \textbf{67}  &  277\\ 
Oblateness  &  1  &  \textbf{31}  &  62  &  9  &  \textbf{41}  &  92  &  10  &  \textbf{38}  &  76\\ 
Virial $\alpha$  &  3  &  \textbf{35}  &  254  &  14  &  \textbf{47}  &  141  &  8  &  \textbf{42}  &  144\\ 
Internal $\mathcal{M}$  &  1  &  \textbf{14}  &  50  &  5  &  \textbf{14}  &  32  &  2  &  \textbf{22}  &  65\\ 
Total $\mathcal{M}$  &  0  &  \textbf{6}  &  18  &  1  &  \textbf{9}  &  17  &  2  &  \textbf{13}  &  40\\ 
Alfven $\mathcal{M}$  &  2  &  \textbf{27}  &  104  &  4  &  \textbf{17}  &  38  &  7  &  \textbf{30}  &  72\\ 
$\left|\vec{j} \right|$  &  3  &  \textbf{52}  &  232  &  5  &  \textbf{59}  &  154  &  10  &  \textbf{80}  &  242\\ 
(Max($\vec{j}$)-Min($\vec{j}$))/Mean($\vec{j}$)  &  0  &  \textbf{24}  &  52  &  8  &  \textbf{22}  &  64  &  7  &  \textbf{31}  &  61\\ 
$\left|\vec{B} \right|$  &  1  &  \textbf{17}  &  48  &  5  &  \textbf{15}  &  45  &  4  &  \textbf{25}  &  60\\ 
(Max($\vec{B}$)-Min($\vec{B}$))/Mean($\vec{B}$)  &  1  &  \textbf{25}  &  71  &  7  &  \textbf{22}  &  43  &  8  &  \textbf{30}  &  50\\ 
Plasma $\beta$  &  2  &  \textbf{36}  &  154  &  3  &  \textbf{26}  &  67  &  15  &  \textbf{43}  &  156\\ 
\hline 
$\Delta$  &  19  &  \textbf{85}  &  270  &  31  &  \textbf{80}  &  166  &  34  &  \textbf{95}  &  272\\ 
\end{tabular}

}
\end{table*}
}

\subsection{Isolated, starless cores}\label{ss:isolated}

\begin{figure}
\centering
\includegraphics[width=\columnwidth]{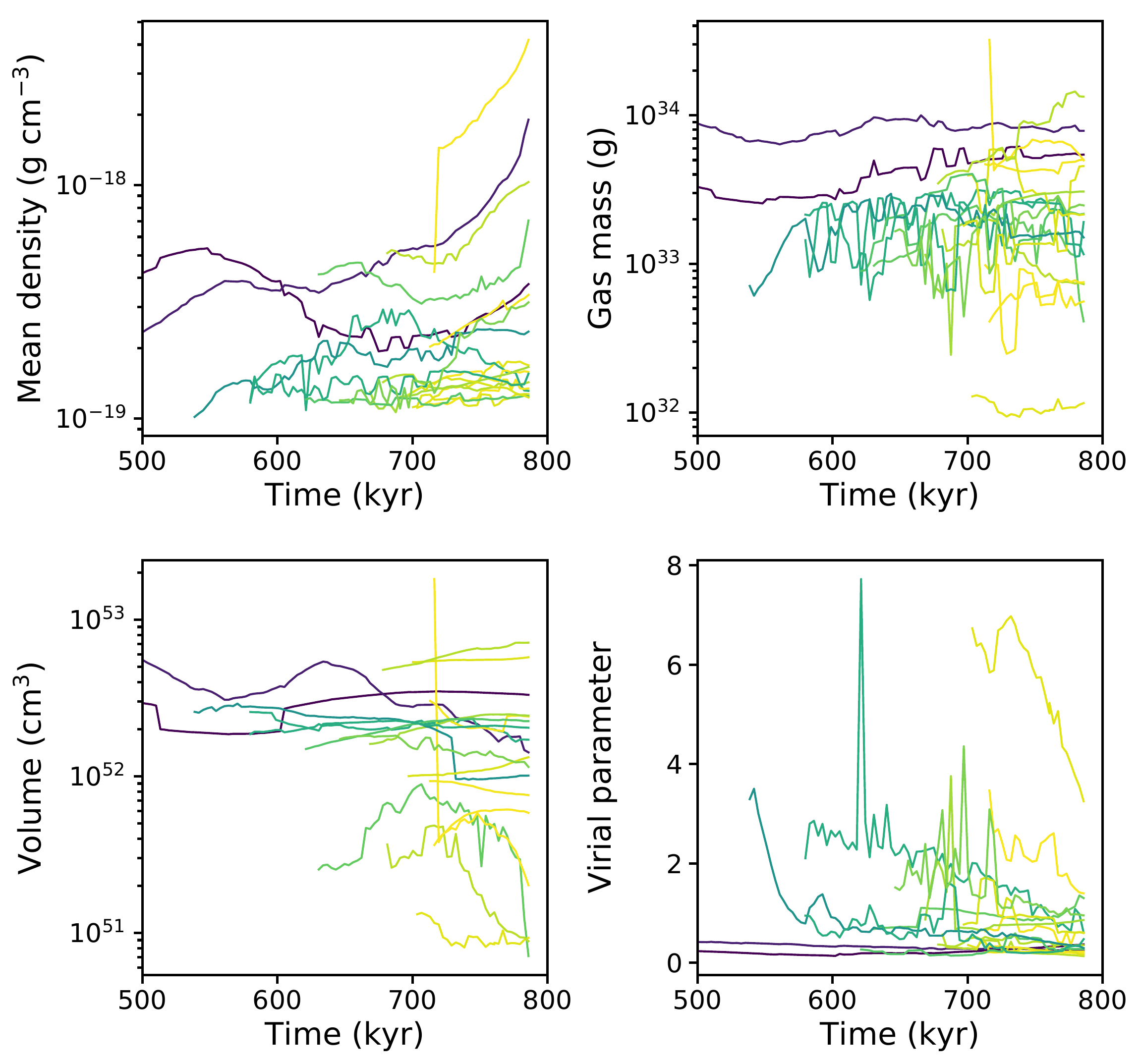} % plot_isolated_starless.py
\caption{ Individual core evolution for the long-lived, isolated, starless cores in our simulation. We show mean density, gas mass, volume, and virial parameter in the four panels.   Most cores show stochastic variation on the order of a factor of a few over the course of their lifetimes. 
\label{f:isolated}}
\end{figure}

Naively, we might expect the long-lived, isolated, starless cores  in our simulations to show the least variation.  \changed{Observed lower} density cores can have lifetimes of upwards of 1\,Myr, which suggests that these cores should vary slowly over their lifetime if they are free from external influence \citep{Andre2014}.  Thus, we analyze these isolated cores separately; Table~\ref{tab:iso} shows the coefficient of variance for \changed{the 18} isolated (one path in the family), long lived ($t>75$kyr), starless leaves.  While some vary by only a few percent, other paths have variability of CV$>40$\% for the 15 parameters in the table. This indicates that these leaves have large computed variability, in contrast to our naive expectation.  The individual leaf evolution tracks are shown in Figure~\ref{f:isolated}, where we plot mean density, mass, volume, and virial parameter.
Most leaves show fairly large stochastic variations in individual quantities on short timescales. These variations are commonly due to changes in the physical structures included in the dendrogram leaf rather than significant physical evolution. However, a few of our leaves (namely, those plotted in purple), show relatively quiescent evolution over their full lifetime, which are akin to the structures identified in \cite{Chen2019}.  These \changed{quiescent} cores will contribute to the statistics of core property distributions while not participating in the star formation process, thereby confusing the mapping of the core mass function to the initial mass function of stars \citep[e.g.][]{Offner2014}.
{
\begin{table}
\centering
\caption{Coefficient of variance range for isolated, long-lived, starless cores \label{tab:iso}}
\begin{tabular}{lccc} 
Quantity & Min (\%) & Mean (\%) & Max (\%)\\
\hline
Total mass  &  10  &  \textbf{30}  &  53\\ 
Gas mass  &  10  &  \textbf{30}  &  53\\ 
Volume  &  2  &  \textbf{18}  &  61\\ 
Size  &  6  &  \textbf{45}  &  277\\ 
Mean density  &  3  &  \textbf{17}  &  54\\ 
%Pressure  &  3  &  \textbf{17}  &  54\\ 
Oblateness  &  9  &  \textbf{33}  &  89\\ 
Virial $\alpha$  &  14  &  \textbf{35}  &  75\\ 
Internal $\mathcal{M}$  &  2  &  \textbf{12}  &  27\\ 
Total $\mathcal{M}$  &  1  &  \textbf{8}  &  20\\ 
Alfven $\mathcal{M}$  &  4  &  \textbf{15}  &  30\\ 
$\left|\vec{j} \right|$  &  5  &  \textbf{48}  &  154\\ 
(Max($\vec{j}$)-Min($\vec{j}$))/Mean($\vec{j}$)  &  8  &  \textbf{28}  &  64\\ 
$\left|\vec{B} \right|$  &  5  &  \textbf{14}  &  56\\ 
(Max($\vec{B}$)-Min($\vec{B}$))/Mean($\vec{B}$)  &  7  &  \textbf{21}  &  45\\ 
Plasma $\beta$  &  6  &  \textbf{21}  &  67\\ 
\end{tabular}

\end{table}
}

\subsection{Virial evolution of cores}\label{ss:virial}

\begin{figure}
\centering
\includegraphics[width=\columnwidth]{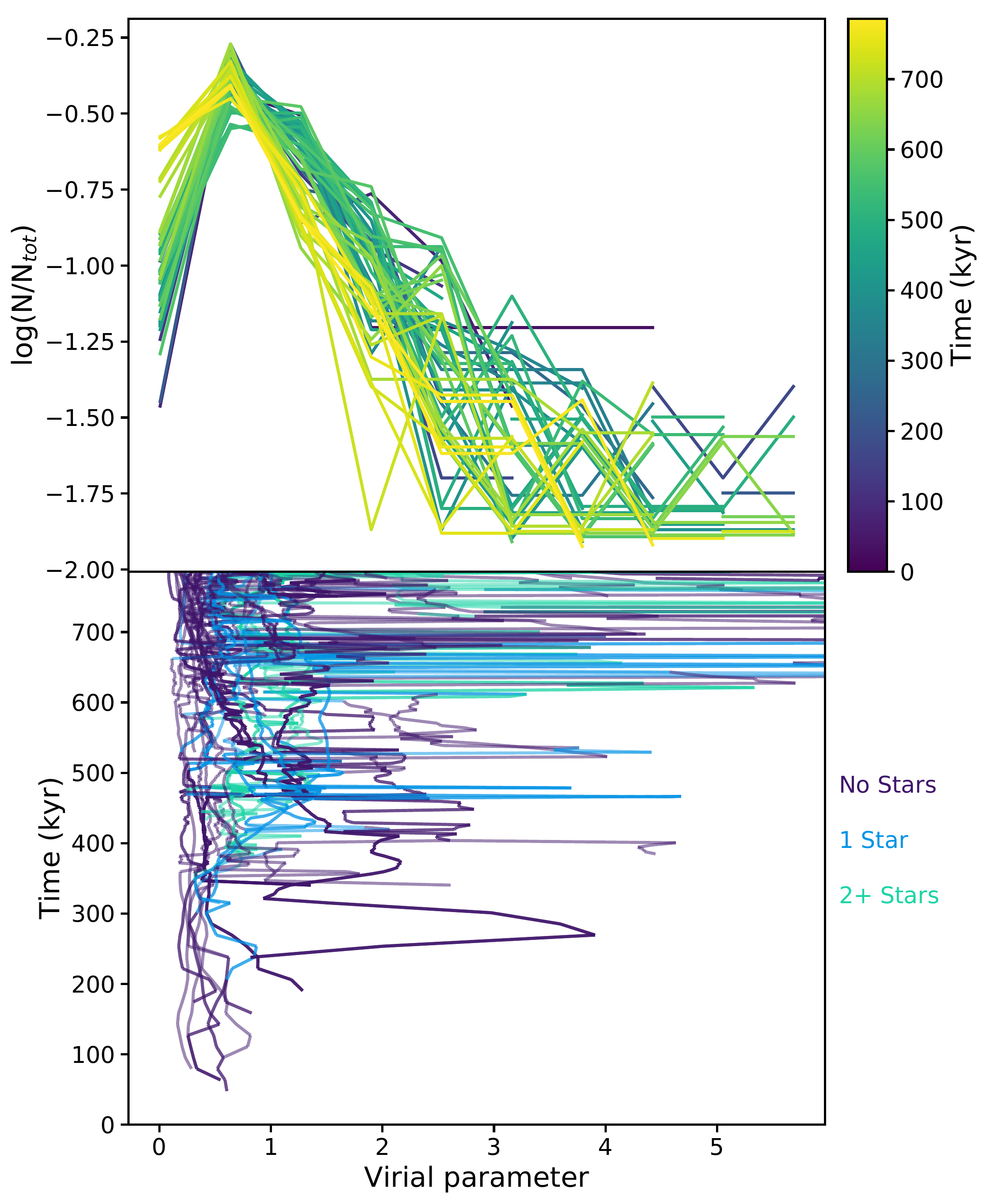}% plot_cmf_withlines_andtime.py
\caption{ Core property distribution vs.~individual core evolution of the virial parameter.  The top panel shows the distribution of the virial parameter of the cores across time.  The bottom panel shows the virial evolution of a subset of reconstructed paths through time.  Coloring is the same as Figure~\ref{f:dist}. 
\label{f:dist_vir}}
\end{figure}

Despite the wide variability in the time evolution of other core properties identified with dendrograms, the virial evolution of leaves does trend in the expected direction of lower virial numbers with time \changed{as seen in the lower panel of Figure~\ref{f:dist_vir}}.  This fits the classic view of star formation where a { magnetized} core undergoing global gravitational contraction will eventually become supercritical and collapse to form a star \citep{Mouschovias1976}. However, our core tracking algorithm does not find strong evidence that a leaf with low virial $\alpha$ will form a star as shown in Figure~\ref{f:dist_vir}. Note that we are using the simplified gravitational $\alpha$ \citep[which is frequently used in observations, e.g.,][]{Kirk2017} and not computing the full virial $\alpha$ that includes boundary terms. Most leaves \changed{($>70\%$)}, especially those hosting sink particles, do finish the simulations with $\alpha < 2$. However, a substantial population of the long-lived, starless leaves have $\alpha < 2$ as well. Many of these low-$\alpha$ leaves persist for longer than a local free-fall time (a few hundred kyr) without forming a star.  Thus, virial $\alpha$ is not necessarily the best predictor of future star formation; other physics, such as pressure or magnetic support, are important factors in the global evolution of a core.

\subsection{Short-lived overdensities}\label{ss:noise}

\begin{figure}
\centering
\includegraphics[width=\columnwidth]{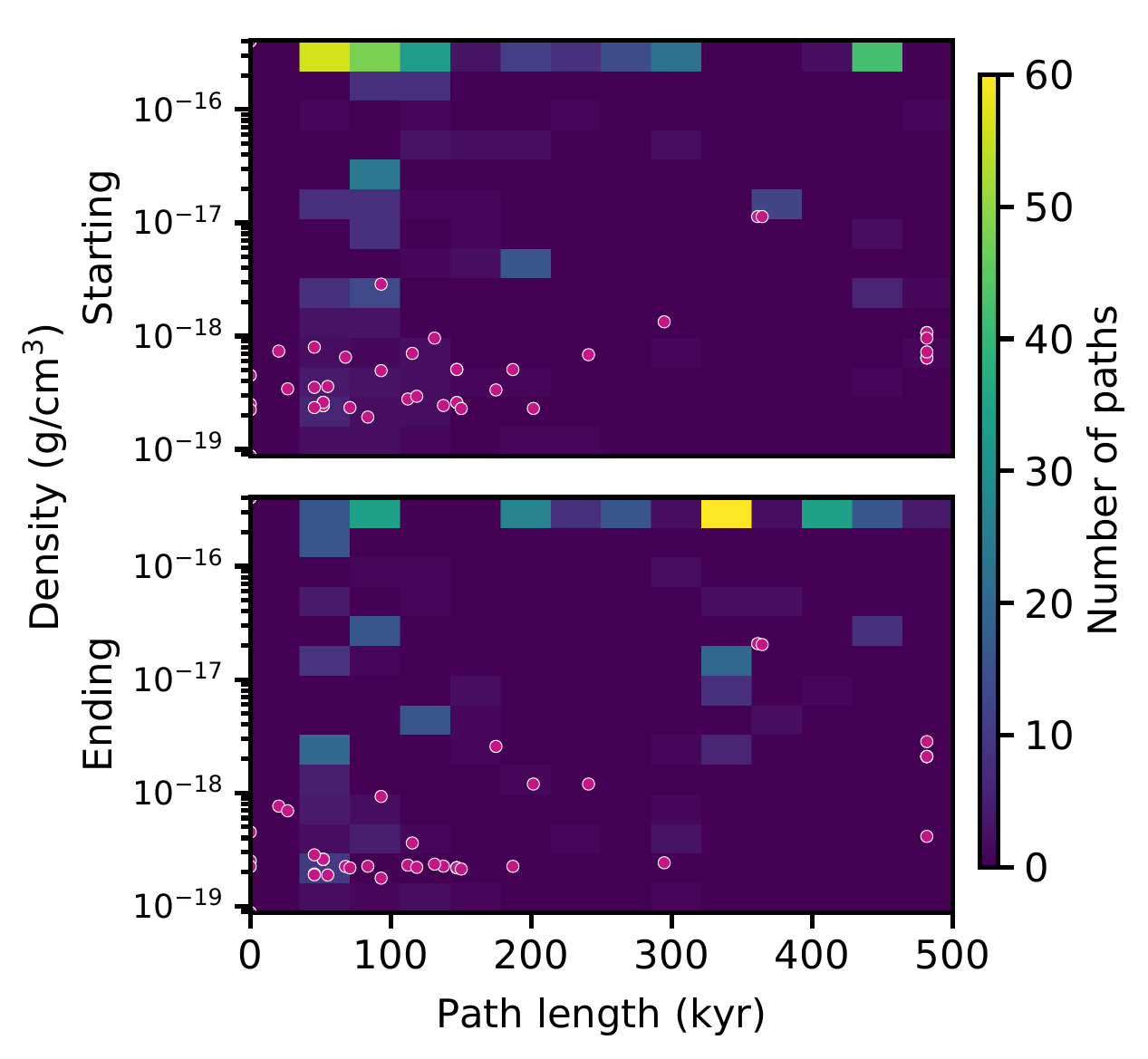} %plot_forward_params.py
\caption{ Starting and ending densities of isolated objects.  The background histogram shows the starting and ending peak densities of all paths (top and bottom, respectively) plotted against the length of the path.  The purple points denote isolated paths.  There is a substantial population of low-density leaves with lifetimes less than 150 kyr  (25\% of all path families) that are temporary, non-gravitating overdensities that  disperse and fall below the dendrogram floor. 
\label{f:dens_2dhist}}
\end{figure}

We observe a population of short-lived, low-density peaks arising from turbulent flows that contribute a level of ``noise'' to the interpretation of long-term core evolution, \changed{since they do not go on to collapse and form protostars.}  These overdensities account for about 25\% of path families identified when tracing paths forward from an intermediate timestep.  We identify this temporary population of \changed{``imposter cores''} as having lifetimes less than 200 kyr and densities less than $1\times10^{-18}$\gcm, as can be seen in Figure~\ref{f:dens_2dhist}.   The majority of isolated paths occupy a much lower density and shorter lifetime than the general path population.  Most of the paths in the low-density and short lifetime region also show the trend that the maximum density \changed{(which is typically also less than $1\times10^{-18}$\gcm) } is higher than the last identified density \changed{(the ending density of the path)}, suggesting that these leaves are physically temporary overdensities that decay below the threshold density required to be identified in the dendrogram.  The free-fall time of these overdensities is about 100 kyr; because the free-fall time is roughly equivalent to the overdensity lifetime, these objects are not dominated by gravitational collapse. 

The presence of this substantial population of imposter cores could introduce a bias in the instantaneous core mass function.  These overdensities have gas masses of order one solar mass and sizes of roughly a tenth of a parsec, which is similar to masses and sizes of observed cores and may therefore masquerade as pre-star-forming cores.  Thus, at any given time, roughly 15-25\% of cores identified in a region may be from this temporary population.   \changed{We performed a two-sample Kolmogorov-Smirnov test on the computed core properties of imposter cores compared to all other cores at the same time: we could not distinguish differences in the distributions of any parameter except density.  For instance, gas mass and virial parameter both have a p-value of 3\% and K-S statistic of 0.4; if the p-value is high (our preferred cutoff is more than 1\%) and the K-S statistic is low (our preferred cutoff is less than 0.6), we cannot claim that the two samples are drawn from different populations. Indeed, p-values for our computed parameters except for density are typically above 3-5\% and K-S statistics are typically less than 0.5.  Thus, imposter cores are not easily separated from any other population of cores, so they will complicate the correlation of core and stellar properties. These temporary overdensities are explored in more detail in Chen (in prep.). }

\subsection{Core mass function}\label{ss:cmf}

\begin{figure}
\centering
\includegraphics[width=\columnwidth]{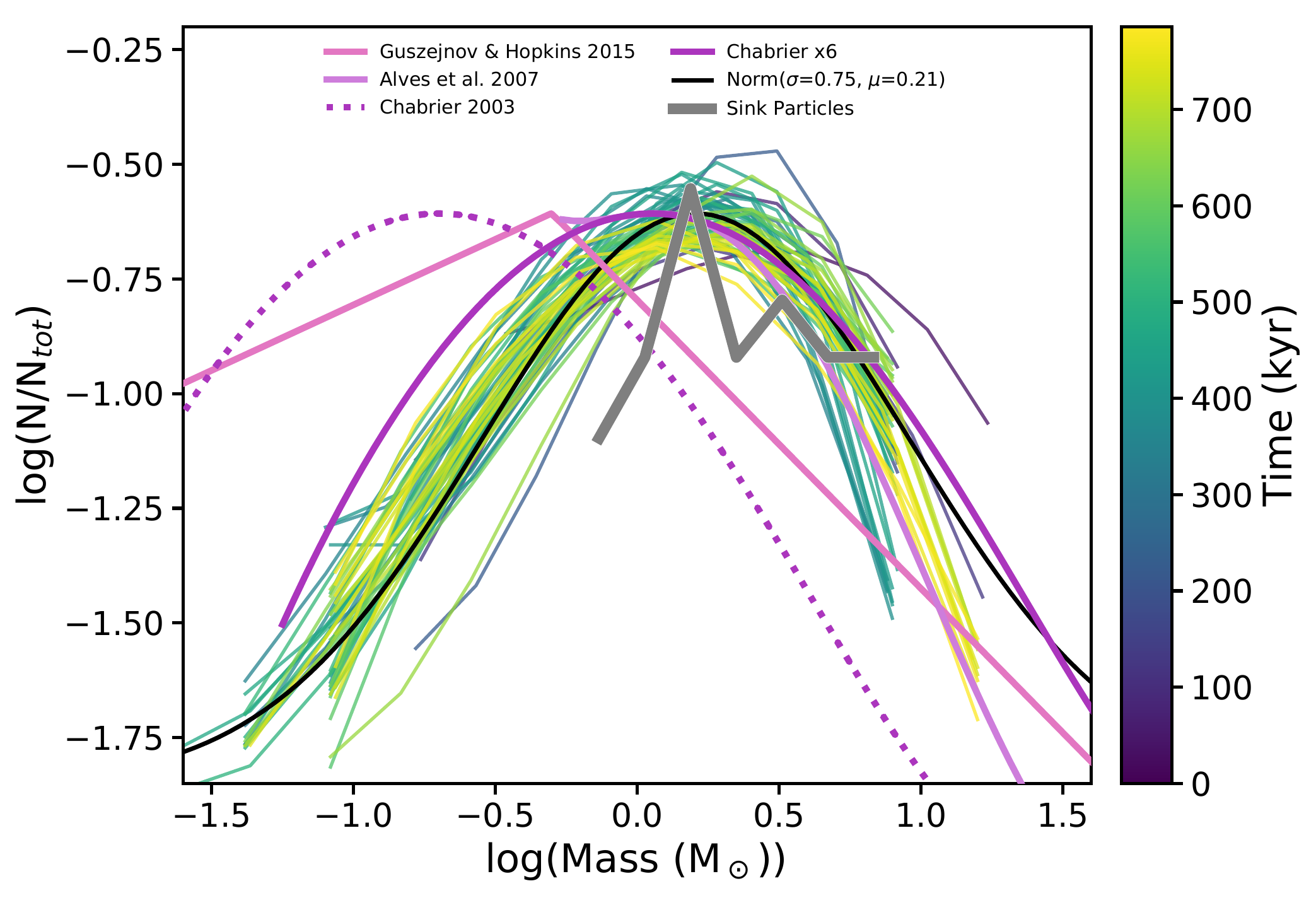} %plot_cmf_withlines.py
\caption{ The core mass function (CMF) across time in our simulations. The purple to yellow color scale show our normalized core masses for a selection of timesteps.  We plot the total core mass, which includes mass from both gas and sink particles.  The mass function of our sink particles (which are equivalent to a protostar and compact disk) is shown in the thick gray line.  We also show the fiducial CMF from \protect\cite{Guszejnov2015}, the observed CMF from \protect\cite{Alves2007}, and the initial mass function (IMF) inferred from observations from \protect\cite{Chabrier2003}.  We plot a log normal in black.
\label{f:cmf}}
\end{figure}

We measure the core mass function (CMF) of our simulation. We show our CMF through time in Figure~\ref{f:cmf} together with mass functions from the literature.  We compute the total core mass, which includes mass from both gas and sink particles.   We compare our CMF to the fiducial CMF from \cite{Guszejnov2015}, the observed CMF from \cite{Alves2007}, the initial mass function (IMF) inferred from observations from \cite{Chabrier2003}, and a log normal \changed{distribution}.

As is seen in the figure, our CMF is relatively invariant through time.  Our peak mass is constant at around 1.4\msun with a range from about $0.3-10$\msun.  There is a small \changed{perceived} bias towards higher masses at earlier times, which is a byproduct of \changed{low number statistics and} the lack of significantly refined structure in our simulations shortly after gravity is turned on.  The constant nature of the CMF is likely due to two effects. The dendrogram introduces new leaves when temporary overdensities are significant enough to warrant leaf creation, leading to the transient population of low-mass ``cores'' discussed above that balances the small physical growth of persistent cores and the algorithmic fragmentation of more massive leaves into smaller structures.  The trend of nearly constant CMF across time in a singular environment is seen in other simulations, such as \cite{Cunningham2018}, where cores mass distributions do not show significant mass evolution after formation.
As described above, any given leaf can occupy a wide range of the total mass space as we track it through time, but the ensemble of leaves maintains a constant distribution in time. Thus, the CMF derived from a dendrogram population does not necessarily correlate with the final IMF of the region; the stochastic nature of leaf mass evolution makes it very difficult to compute a relationship between a core mass at any snapshot and the resultant stellar mass. 
The CMF in our simulations agrees well with the observed CMF from \cite{Alves2007}, with similar mass peaks.  We also find a good agreement with a \cite{Chabrier2003} system IMF scaled by a factor of 6.  We do not, however, create the population of low mass cores of the \cite{Guszejnov2015} model.  We also do not create the population of low mass sinks particles seen by other simulations such as \cite{Bate2012} despite our ability to create sink particles with masses much less than 1\msun (although this is expected due to the coarse spatial resolution and lack of feedback).  \changed{In this work, we do not intend to explain the evolution of the \changed{CMF to the} IMF; instead, we simply \changed{aim} to show that our method produces cores that are broadly comparable to observations.}

\begin{figure}
\centering
\includegraphics[width=\columnwidth]{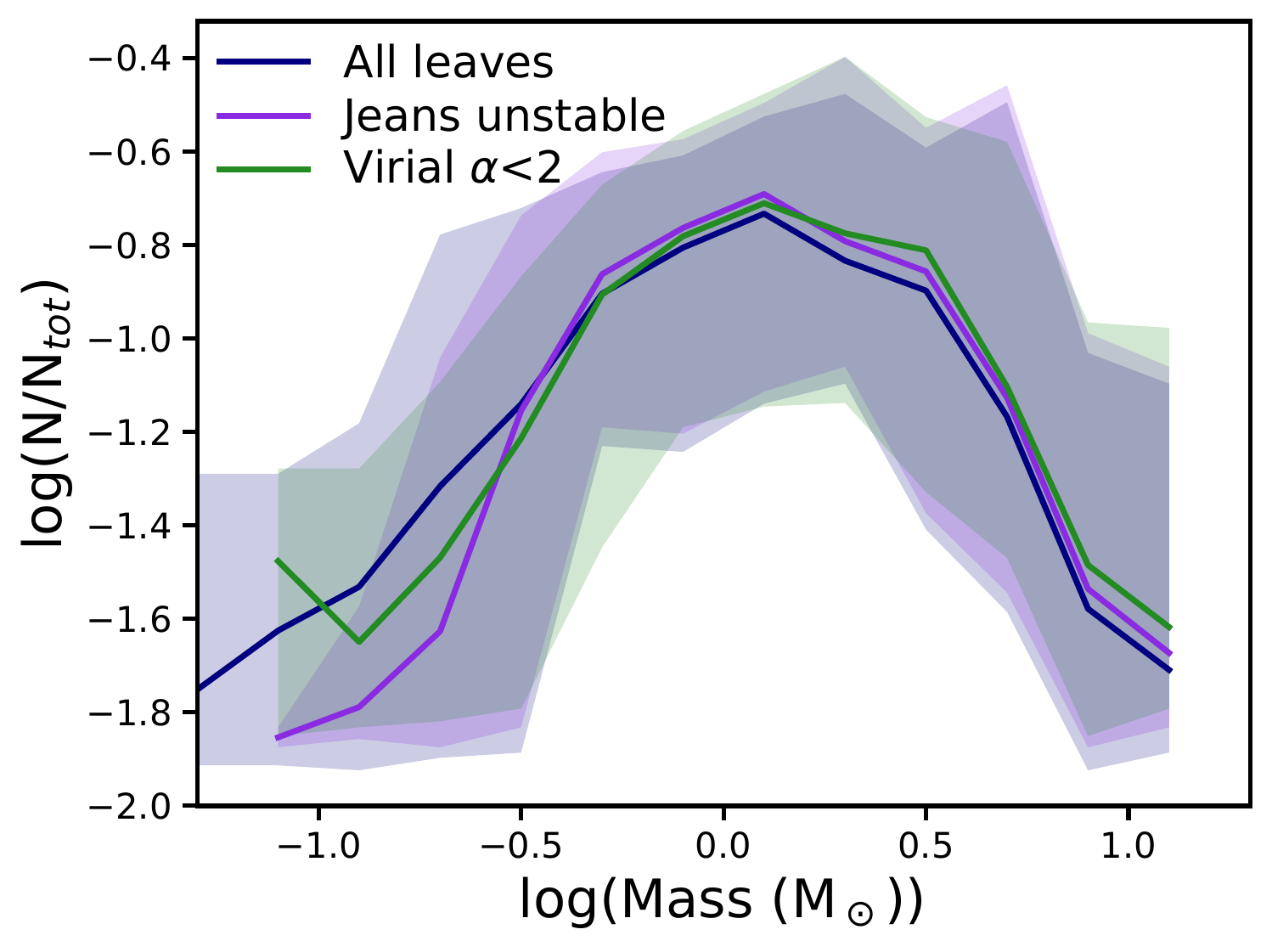} %plot_cmf_compare.py
\caption{ \changed{The core mass function for different core definitions.  Solid lines show an average CMF across time, while the shaded regions show the minimum and maximum bins across time.  Blue shows the CMF of all leaves in the simulation (which is what is shown in Figure~\ref{f:cmf}). The other colors show core selections that might be more physically motivated: purple shows cores that have masses greater than the local Jeans mass, and green shows cores that have virial $\alpha<2$.  All mean CMFs overlap except at the lowest mass end, where some of the low mass cores don't satisfy the stricter Jeans or Virial criteria.}
\label{f:cmf_compare}}
\end{figure}

\changed{We compare the CMF derived from the leaf gas mass to the CMF derived from more observationally-motivated core definitions including the leaves that eventually form stars (equivalent to prestellar and protostellar cores), leaves with $\alpha<2$, and leaves that are Jeans unstable for their mean density.  All of these different populations produce quantitatively similar results, as shown in Figure~\ref{f:cmf_compare}.  The CMFs all have peaks slightly higher than 1\msun, a spread of about two orders of magnitude, and are invariant in time.  We therefore conclude that any structures not involved in the traditional star-formation process (e.g., \changed{transient} overdensities) have little impact on the derived CMF.  Computing the CMF based on different properties in observations also produce similarly invariant CMFs \citep[e.g.][]{Sokol2019}. }

The highly variable masses of identified leaves through time means that we cannot infer the IMF by looking at a population of cores identified with dendrograms at a given time snapshot.  While there may be an underlying physical evolution of star-forming cores, the instantaneous properties of a region identified by dendrogram cannot be assured to correlate with that evolution. 

\section{Results: Insights from methodology}\label{s:dendros}\label{ss:caveats}

It is imperative to understand the impact of the dendrogram algorithm on  cores identified in both simulations and observations due to the algorithm's wide popularity in the literature. In this section, we discuss the insights into the use of dendrograms gained from this work.

We have used dendrograms to identify dense structures in our simulations of a star-forming region, but we have also demonstrated a limitation of dendrograms: because dendrograms identify  \emph{relative} variations in structure, leaf structure may vary significantly between timesteps due to small variations in the local density structures.  Dendrograms are built beginning from the maximum value, so any variations in that maximum may cascade into substantial changes in the resultant dendrogram architecture. 

\begin{figure}
\centering
\includegraphics[width=1\columnwidth]{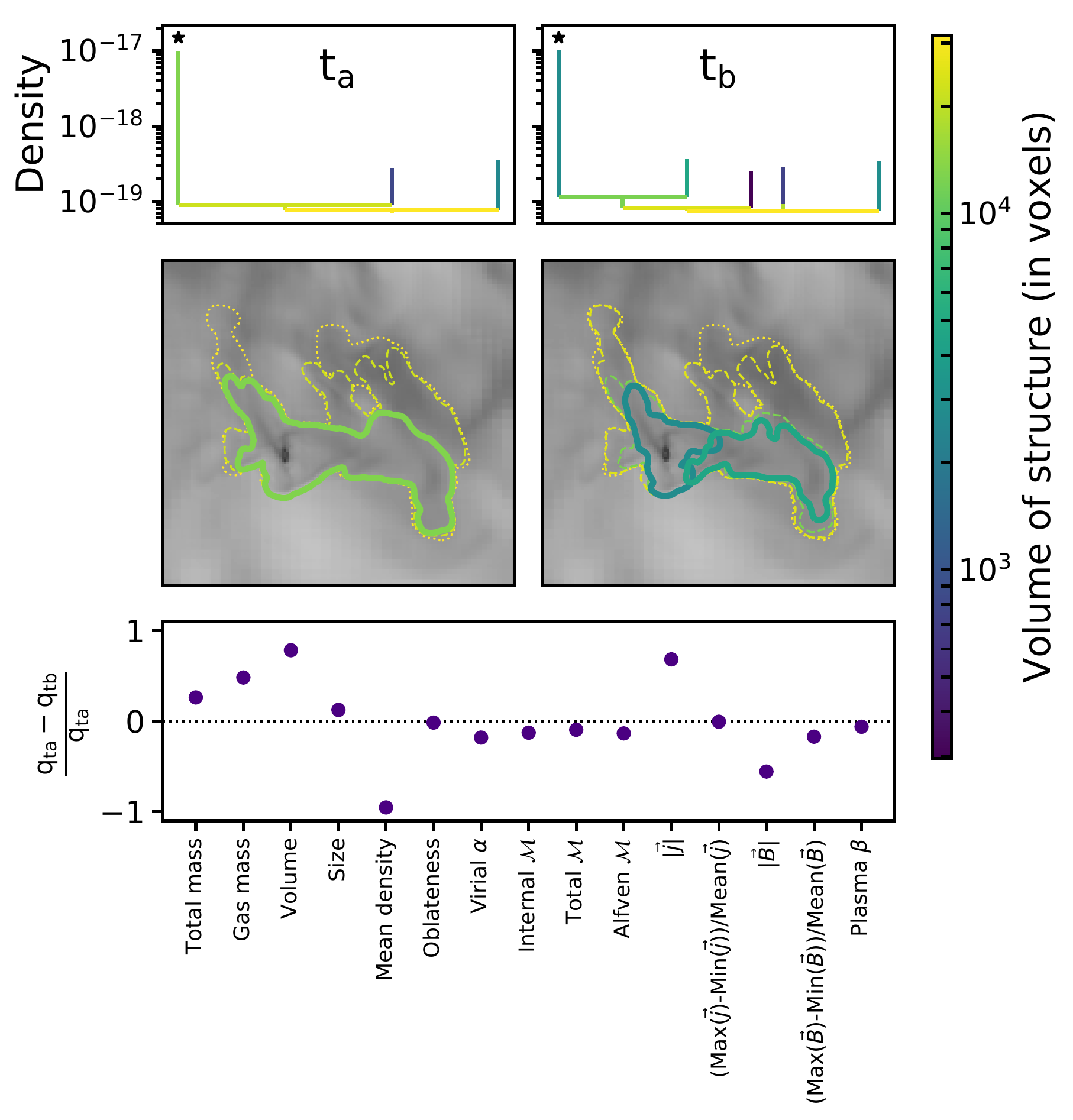}%plot_tree_height_quants.py
\caption{ The dendrograms of a small volume in consecutive timesteps.  The left and right columns depict different timesteps.  The upper panels show the dendrogram structure colored by leaf volume. The starred leaf in each panel is the leaf containing the dominant overdensity in the middle panels.  The middle panels show the leaf contours over a grayscale density projection of the simulation.  Dotted contours show the trunk, dashed contours show branches, and solid contours show leaves.  Despite very little physical evolution between timesteps, the dendrogram identifies different tree structure, leading to significantly different leaf morphologies. The bottom panel shows the impact of the different leaf structure on computed leaf properties for the leaf starred in the upper panels. Critical quantities such as mass show significant differences between the two times that can only be attributed to the redefinition of the leaf contours.
\label{f:compare_consecutive}}
\end{figure}

An example of this phenomenon is shown in Figure~\ref{f:compare_consecutive}.  The left and right columns depict neighboring timesteps.   Despite very little physical evolution between timesteps (a $\unsim5$\% change in the peak density), the dendrogram identifies leaf structure quite differently. This translates to a nearly order of magnitude variation in the volume of the sink-hosting leaf and substantial variation in the computed properties of the leaf.  

The two leaves in the right panel that are part of the sink-hosting sub-tree (the two left-most leaves in the dendrogram) are not physically interacting over the course of the simulation.  They are simply nearby overdensities.  However, because of the variations in the dendrogram structure, our algorithm identifies these two leaves as belonging to the same path family. Thus, one of the major failings of tracking overdensitites identified via dendrogram through a simulation to study core evolution is that it becomes difficult to disentangle physical evolution from ``algorithmic'' \changed{artifacts}.  In other words, there is no easy, automated way to differentiate between physical structure change and dendrogram structure change.

\begin{figure}
\centering
\includegraphics[width=0.7\columnwidth]{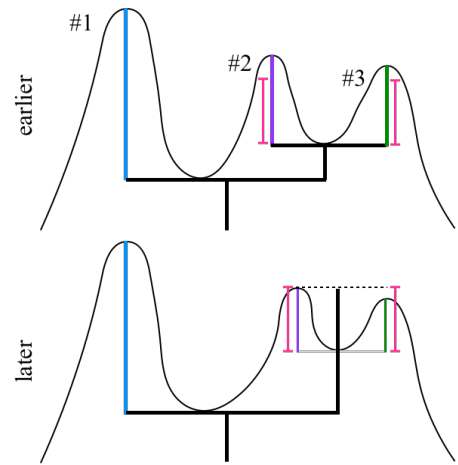}
\caption{ Cartoon explanation of the origin of algorithmic structure variation in time. Structures 1 (blue), 2 (purple), and 3 (green) are shown at two consecutive times.  The physical properties of the structures (peak and width) don't change between the top (earlier time) and bottom (later time).  However, their relative locations with respect to one another do change: \changed{structures 2 and 3 move closer to one another, thereby increasing the density of the saddle point between them.} This causes the nodes (horizontal black lines) to be at different heights at the two different times. At the earlier time, both structures 2 and 3 exceed the density increase criterion (indicated by the pink vertical lines), while at the later times, the individual density peaks of 2 and 3 are not significant enough to allow for substructure to be identified.
\label{f:dendro_explain}}
\end{figure}

The change in consecutive dendrograms arises because of small variation in the relative properties of structures (typically intermediate density structure).  Figure~\ref{f:dendro_explain} aims to illustrate the issue.  Structures 1, 2, and 3 are shown at two consecutive times.  The physical properties of the structures (peak and width, in this cartoon) don't change between the top (earlier time) and bottom (later time).  However, their relative locations with respect to one another do change.  \changed{The structures have moved closer to one another and therefore the saddle point between them has become shallower}. This causes the nodes (horizontal lines) to be at different heights at the two different times. At the earlier time, the node is low enough that both structures 2 and 3 exceed the density increase criterion (indicated by the pink vertical lines), while at the later times, the node is at a high enough density that the individual density peaks are not significant enough to allow substructure to be identified.

To further explain the example presented from this work, the second leaf in the right panel of Figure~\ref{f:compare_consecutive} is just above the density refinement criterion at its physical location in that timestep.  However, the peak density in that region drops by 4\% in the left panel, which then leads to the overdensity not being quite ``peaky'' enough to satisfy the density refinement when compared with the maximum peak.  This is not a problem unique to our density refinement criterion: any density refinement chosen will exhibit these \changed{artifacts} to some degree due to the relative nature of dendrogram structure identification.  Even observationally, these issues may be seen: differences in resolution or noise levels in observations of the same region may lead to changes in the computed hierarchy. \changed{Any variation between consecutive observations in the region around a peak dendrogram can lead to variations in the contour drawn by the dendrogram.}

\section{Discussion}\label{s:discussion}

\subsection{Interpreting the IMF from the CMF}\label{ss:imf_cmf}
\begin{figure}
\centering
\includegraphics[width=\columnwidth]{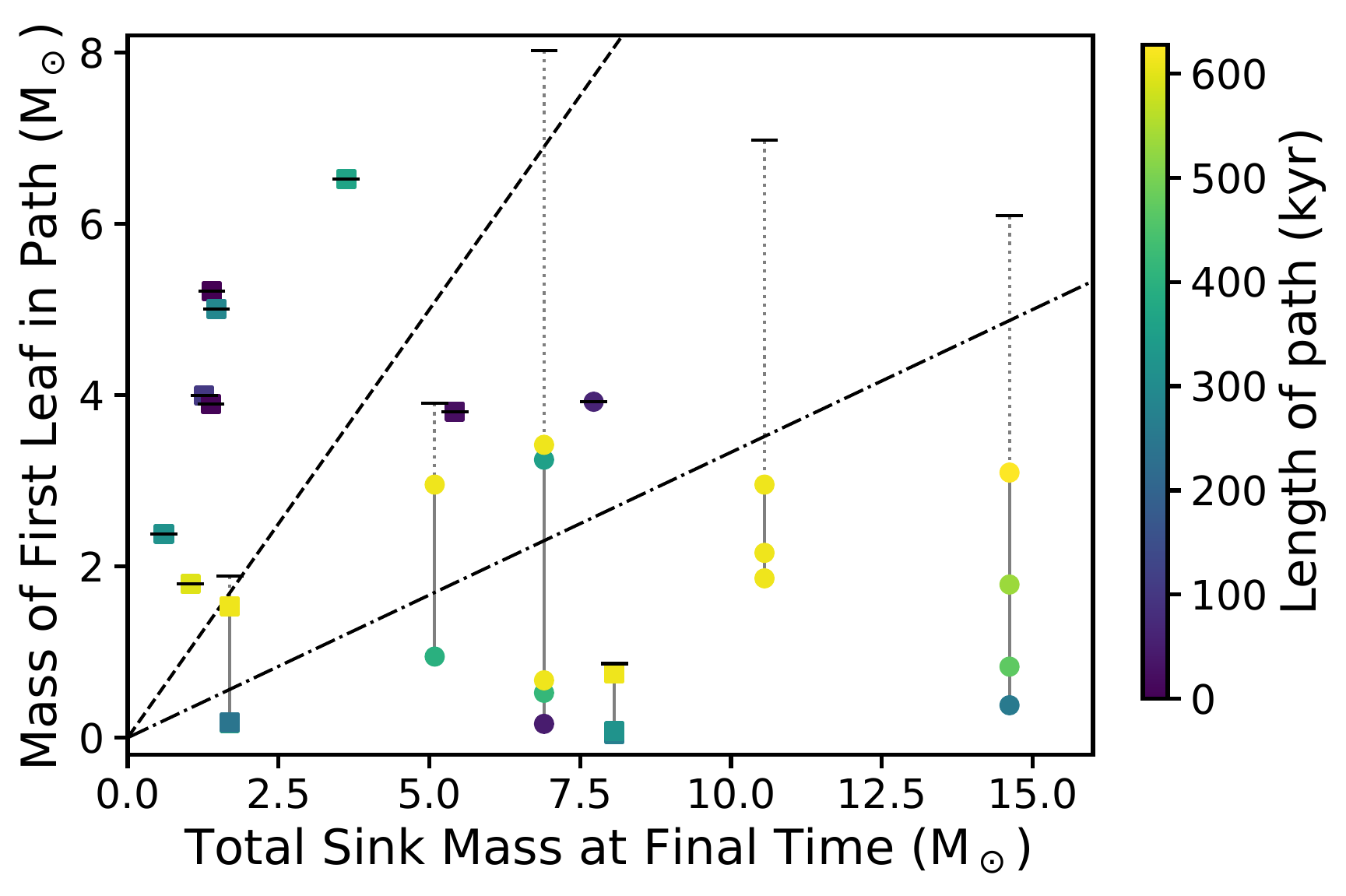} %plot_msink_v_mgas_dosum.py
\caption{ Sink mass at the final time plotted against leaf mass.  Squares show leaves containing one star at the final time, while circles show leaves containing multiple stars.  Points are colored by the length of the path, and we have truncated the paths to have an earliest age of 250 kyr after the gravity turned on.  Systems that consist of multiple paths are connected by a vertical line.  The sum of the leaf masses is indicated by the horizontal marker.  The dashed line shows the 1:1 correlation, while the dot-dashed line shows the \changed{trend if the sink mass is reduced by the factor of 3 efficiency arising from the lack of protostellar feedback in the simulations.}  Most systems fall above the efficiency factor line, but including the sum of leaf masses is critical for the most massive systems. 
\label{f:imfvcmf}}
\end{figure}

One natural question we can ask in this work is how the instantaneous core masses correlate with the stars they form.  We plot this in  Figure~\ref{f:imfvcmf}, where we show the total sink mass at the end against the initial masses of leaves that merge into the final overdensity.  First, there is a wide array of scatter in the initial leaf masses that doesn't correlate well with the final sink mass (although some of this scatter may result from the lack of feedback in our simulations). This observation is consistent with other works for low to intermediate-mass stars such as \cite{Smith2009a} and \cite{Mairs2014}. Second, the \emph{sum} of all component leaf masses seems to be a very important consideration, especially when considering the growth of systems containing multiples.    For the most massive multiple systems, any individual leaf does not fall above the factor of three efficiency threshold \changed{for gas conversion into protostars} commonly used in the literature \citep[e.g.][]{Alves2007} which likely arises due to protostellar outflow feedback \changed{and is not accounted for in the mass accretion of sink particles in this simulation} \citep{Offner2014,Offner2017}, but the sum of the leaf masses puts the systems into a comparable space as all other systems.  It is important to note that the systems containing multiple stars do not form by multiple leaves containing single stars merging together; rather, the bound multiple forms in one leaf, and the accretion of gas overdensities may help trigger the formation of new stars.  This result is similar to that seen in \cite{Padoan2019}, who see little correlation between core mass (or even extended mass around a core) and stellar mass for high-mass stars. \changed{This observation supports binary formation models such as turbulent fragmentation in a single core \citep[e.g.,][]{Offner2010} or disk fragmentation \citep[e.g.,][]{Kratter2010} as opposed to dynamical capture \citep[e.g.,][]{Bate2003}.}

\subsection{Other ways to identify cores}\label{ss:identify}
There are a few ways one may attempt to overcome the limitations of core identification and comparison using density dendrograms. We discuss the \changed{advantages} and limitations of these ideas.

One could try to overcome the impositions of the dendrogram algorithm itself.  Custom merging strategies are possible in \texttt{astrodendro}; a ``pruning'' strategy will allow peaks near the density refinement criterion to remain more stable and thereby overcome some of the relative structure \changed{variation} shown in Section~\ref{ss:caveats}.  However, this solution {is} subject to human bias due to the addition of another tunable parameter.   It is unclear how to create a custom merger strategy in a way that is agnostic to the human-desired structure without introducing more bias.

In principle, one could also create contours at absolute density levels instead of relying on a relative measure. By using an absolute density contour, the leaf structure should slowly vary from timestep to timestep and may therefore better identify bound cores.  One could then create a hierarchical structure tree that is similar to a dendrogram, but the nature of hierarchy would be more difficult to determine due to the fact that many density peaks would be broken into a single nested hierarchy. Additionally, this type of hierarchy would destroy the physical utility of dendrograms in studying the relation of physical structures in a region and again relies on an arbitrary density threshold, which we advocate against.

Core identification \changed{in simulations} might also be better done in 2-D synthetic observation space instead of 3-D density grids \changed{because there will be fewer variations in integrated intensity between timesteps}. 
However, this method is best suited to isolated cores and may suffer from false over-densities in emission created by chance alignments \citep[e.g.,][]{Beaumont2014}. 

Finally, density may not even be the best tracer of star-forming cores as cores are highly dynamic and will not be defined by the same density contour across time. It is likely that a more physically motivated property such as virial parameter, velocity dispersion, or gravitational potential could be a better quantity with which to build hierarchical structures \citep[see, for instance][]{Mao2019}.  
These properties should be less variable across time and should therefore provide a more stable core identification. However, these quantities are more difficult to measure observationally and will make comparisons between simulations and observations harder. 

In simulations, one can also include tracer particles that will trace the evolution of gas in a core identification-independent way.  However, interpreting that evolution is non-trivial. \cite{Smith2009a} and others find that a non-negligible fraction of tracer particles in a bound gas clump will accrete onto a sink particle outside of that bound clump.  Indeed, they find that most of the mass in a sink particle can be accreted from outside its nascent core.  \changed{Thus, the meaning of a core in this context becomes even less apparent, as the star may contain gas from all around the molecular cloud.}

\changed{One could employ alternate core identification algorithms used in the field.  All of these other methods (\textsc{clumpfind}, FellWalker, GaussClumps, etc.), would suffer similar issues because they all fundamentally rely on the relative positions of peaks to determine structure. Different algorithms might have different sensitivity to the less dense material surrounding density peaks, but all algorithms have some way to combine peaks that are thought to not be independent.  }

Each of the methods discussed above would likely identify the same dense gas structures, but the variations in core identification would still likely lead to changes in computed core properties between methodologies.  There is therefore no unique way to define a core in both simulations and observations using existing methods.

\subsection{Implications of core identification}\label{ss:implications}

This work shows that there is no time-stable density contour with which to define cores.  Because of the dynamic nature of core evolution, a single set of dendrogram parameters will not trace unique core parameters across the entire lifetime of core formation.  
\changed{Additionally, we show that a substantial change in the cloud properties (due to time evolution in this case) are required to see changes in the observed CMF: over $>70\%$ of our simulation snapshots show the same CMF, despite order unity variations in individual cores. Changes in the distribution occur at early times. In the context of our simulations, this is because gravity has had less time to overcome the turbulence in the gas. In real systems, this would correspond to the time when the cloud itself was only weakly bound. This trend suggests that variations in the CMF only coarsely trace the time evolution of a star forming region. Thus variations in CMF from one star forming region to should not be attributed solely to differences in age. } Finally, computing a dendrogram in density or intensity on an observed region introduces an inherent uncertainty in the physical importance of structures identified. Dendrograms have many tunable parameters, so disentangling physical structure from algorithmically imposed structure in an automated fashion is a non-trivial endeavor.  

The large variability in the \changed{computed} core boundaries will likely be less dramatic in observational space due to the integration of the signal along a line of sight.  The lower density material around the edges of our identified leaves will not contribute as much signal, so structures will appear more compact around only the densest part of the core. However, as \cite{Beaumont2013} and citations therein show, \changed{simulation} projections \changed{and observation} are highly subject to projection effects, such as non-physical cores being identified due to a large column of low-density material.  Thus, neither physical nor observational spaces have cores that can be robustly and uniquely defined across all time.

\section{Conclusions}\label{s:conclusions}

In this work, we have presented an algorithm that links dendrogram leaves through time in order to study the evolution of dense cores in MHD simulations.  We aim to understand not only the evolution of the  star-forming gas reservoir in our simulations, but also the manner in which the use of the dendrogram algorithm may bias interpretation of core properties and evolution.  \changed{Ideally, the parameters used for identifying and linking cores are set by the underlying physics. As is shown in this work, we ultimately conclude that there is no robust set of density-based parameters that can  trace coherent cores through time. Additionally, we} find the following:

\begin{enumerate}
\item The distributions of core properties, such as mass, are relatively invariant in time.  The CMF matches well with observed CMF distributions such as \cite{Alves2007} and shifted IMFs such as \cite{Chabrier2003}.  Most property distributions do not show significant trends over long timescales.

\item Individual core histories show large variability ($>40\%$) on short timescales (<100 kyr) that arise from changes in the leaf boundaries. This non-monotonic variability persists across environment (isolated or crowded) and stellar content. Additionally, a leaf history that shows low variability in one parameter will not necessarily show low variability in all parameters.  There are no obvious regular trends in time with the exception of virial parameter (which tends to decrease to $\alpha<2$ as the cores reach the end of the simulation). There is some evidence for long-term evolution in of individual paths traced in other properties that may correspond to physical evolution, but the shorter stochastic variability makes these trends difficult to quantify.

\item  The variability exhibited in our analysis of individual core evolution is \changed{at least partially attributable to} the dendrogram algorithm itself.   Small changes in the relative structure of the density between timesteps can propagate to incredibly large changes in the computed boundaries of structures.  In extreme cases, volumes can change by an order of magnitude between timesteps, leading to nearly 100\% variability in computed core properties.  The sensitivity of the dendrogram to small changes in physical conditions raises concerns about hierarchies identified in both simulations and observations.  For instance, changes in noise or resolution may lead to different hierarchies in the same region.

\item We find a population of short-lived overdensities in each timestep that may serve as a substantial source of ``noise'' for core property distributions in observations.  The overdensities tend to have lower density ($<10^{-18}$\gcm) and lifetimes less than 200 kyr, and they account for $15-25\%$ of identified leaves every timestep.   These overdensities have other properties, such as mass and size, that are comparable to other cores in the simulation  that go on to form stars. 

\item Assessing the full history of cores (including events like mergers) may be important for interpreting the IMF.  We find that, especially for massive multiple star systems, the sum of all initial leaves associated with the multiple is typically required to agree with CMF-IMF scaling assumptions even when inefficiency produced by feedback is taken into account. 

\item There is no time-stable density contour that defines a star-forming core.  The dynamic nature of core formation and evolution means that dendrograms will not trace the same structures across time in a reliable way.  Thus, we urge caution when comparing dendrograms of different ages or environments because differences in the dendrogram may come from the algorithm itself instead of physical changes.
\end{enumerate}

In summary, cores identified with dendrograms are subject to algorithmic limitations that impact the physical interpretation of ``observed'' core boundaries. And yet, understanding the full time evolution of star-forming cores is critical to understanding the end results of star formation, such as  interpreting the relationship (or lack thereof) between the CMF and IMF. We have shown the need for caution when extrapolating instantaneous observations of star-forming cores either forward or backward in time, as cores can have substantial variability both intrinsically and observationally.

\section*{Acknowledgements}
%RAS: Anyone else we should add to the acknowledgements?
We are grateful for the helpful suggestions of our anonymous referee.  We also thank others, including Jaime Pineda and Alyssa Goodman, for insightful discussions during the course of this work. RAS acknowledges support from the National Science Foundation under Grant No. DGE-1143953. ATL, SSRO, and KMK acknowledge support from NASA grant NNX15AT05G. KMK additionally acknowledges support from NASA grant 80NSSC18K0726 and RCSA award ID 26077.  SSRO and HHC acknowledge support from a Cottrell Scholar Award from Research Corporation. The analysis presented herein utilized HPC resources supported by the University of Arizona TRIF, UITS, and RDI and maintained by the UA Research Technologies department. This research made use of \texttt{astrodendro}, a Python package to compute dendrograms of astronomical data, and the software package \yt from \citet{Turk2010}.

\appendix
\bibliography{new_lib}

\section{Computed properties}\label{definitions}

We use typical quantities common in star formation studies.  However, for transparency, we define their algorithmic definitions used in this paper.  Volumes have been calculated using the volume of the leaf on the uniform grid. Quantities are computed on the cell-wise level using the AMR cells identified within the leaf and then summed to a single quantity where indicated.  

The center of mass, which is repeatedly used, is defined as $\mu=\sum m_\textrm{gas} v_\textrm{gas}/\sum m_\textrm{gas}$. Then, iterating over all sinks in a leaf, it is modified as $\mu=(\mu m+v_\textrm{sink} m_\textrm{sink})/(m+m_\textrm{sink})$.
\begin{itemize}

\item Mean density: $\sum m_{\textrm{cell}} \rho_{\textrm{cell}} / \sum m_{\textrm{cell}}$

\item Total mass: $M_{\textrm{gas}} + M_{\textrm{sink}}$

\item Gas mass: $M_{\textrm{gas}}$

\item Volume: $N_{\textrm{uniform cells}}\cdot V_{\textrm{uniform cells}}$

\item Size: 
	\begin{itemize}
    \item Size=$\sqrt{ (x_\textrm{max} -x_\textrm{min})^2 + (y_\textrm{max} -y_\textrm{min})^2+ (z_\textrm{max} -z_\textrm{min})^2} $
	\end{itemize}
    
\item Oblateness:
	\begin{itemize}
    \item $\Delta x=(x_\textrm{max} -x_\textrm{min})$
    \item $\textrm{mag}=\sqrt{ \Delta x^2 + \Delta y^2+ \Delta z^2} $
    \item Oblateness=$\left( \textrm{max}\left[ \Delta x,\Delta y,\Delta z \right] -\textrm{min}\left[ \Delta x,\Delta y,\Delta z \right] \right) / \textrm{mag} $
	\end{itemize}
    
\item Virial Parameter:
\begin{itemize}
\item $\sigma=\sqrt{(\sum m ((v_x-\mu_x)^2 + (v_y-\mu_y)^2 + (v_z-\mu_z)^2))/\sum m}$
\item $R=\sum (m/\rho)^{1/3}$
\item $\alpha= 5 (\sigma/\sqrt{3})^2 R / 3 G M_\textrm{tot}$
\end{itemize}

\item Internal Mach number:
\begin{itemize}
\item $c_s = \sqrt{\gamma P/\rho}$
\item $\mathcal{M}=\sqrt{(\sum m (((v_x-\mu_x)^2 + (v_y-\mu_y)^2 + (v_z-\mu_z)^2)/c_s^2))/\sum m}$
\end{itemize}

\item Core Mach number:
\begin{itemize}
\item $c_s = \sqrt{\gamma P/\rho}$
\item $\mathcal{M}=\sum m (\sqrt{v_x^2 + v_y^2 + v_z^2}/c_s)/\sum m$
\end{itemize}

\item Alfven Mach number:
\begin{itemize}
\item $v_A = \sqrt{B^2/4\pi\rho}$
\item $\mathcal{M}_A=\sqrt{(\sum m (((v_x-\mu_x)^2 + (v_y-\mu_y)^2 + (v_z-\mu_z)^2)/v_A^2))/\sum m}$
\end{itemize}

\item Pressure: $\sum m_{\textrm{cell}} P_{\textrm{cell}} / \sum m_{\textrm{cell}}$

\item Angular momentum magnitude:
\begin{itemize}
\item $r_{\textrm{cor}}=r-\mu_r$
\item $v_{\textrm{cor}}=v-\mu_v$
\item $j=\sum ( mv_{\textrm{cor}} \times r_{\textrm{cor}} ) / \sum m$ 
\item $\textrm{Magnitude}=\sqrt{j_x^2+j_y^2+j_z^2}$
\end{itemize}

\item Angular momentum orientation:
\begin{itemize}
\item $r_\textrm{cor}=r-\mu_r$
\item $v_\textrm{cor}=v-\mu_v$
\item $j=\sum ( mv_{\textrm{cor}} \times r_{\textrm{cor}} ) / \sum m$ 
\item $\textrm{mag}=\sqrt{j_x^2+j_y^2+j_z^2}$
\item $\textrm{Orientation} = (\textrm{max}(j)-\textrm{min}(j))/\textrm{mag}$
\end{itemize}

\item B-field magnitude:
\begin{itemize}
\item $B=\left[ \frac{\sum m B_x}{ \sum m}, \frac{\sum m B_y }{ \sum m},\frac{\sum m B_z }{ \sum m }\right]$
\item $\textrm{Magnitude}=\sqrt{B_x^2+B_y^2+B_z^2}$
\end{itemize}

\item B-field orientation:
\begin{itemize}
\item $B=\left[ \frac{\sum m B_x}{ \sum m}, \frac{\sum m B_y }{ \sum m},\frac{\sum m B_z }{ \sum m }\right]$
\item $\textrm{mag}=\sqrt{B_x^2+B_y^2+B_z^2}$
\item $\textrm{Orientation} = (\textrm{max}(B)-\textrm{min}(B))/\textrm{mag}$
\end{itemize}

\item Plasma $\beta$
\begin{itemize}
\item $\beta_\textrm{cell}=8 \pi P/B^2$
\item $\beta=\sum m_{\textrm{cell}} \beta_\textrm{cell} / \sum m_{\textrm{cell}}$
\end{itemize}

\end{itemize}

% Don't change these lines
\bsp	% typesetting comment
\label{lastpage}
\end{document}